%% file: main.tex
\documentclass[lettersize,journal]{IEEEtran}
\usepackage{amsmath,amsfonts}
\usepackage{array}
\usepackage[caption=false,font=scriptsize,labelfont=sf,textfont=sf]{subfig}
\usepackage{textcomp}
\usepackage{stfloats}
\usepackage{url}
\usepackage{verbatim}
\usepackage{graphicx}
\usepackage{cite}
\usepackage{multirow,booktabs,pifont,siunitx,adjustbox,tikz,xcolor,amssymb,enumitem}
\usepackage[ruled,linesnumbered,noend]{algorithm2e}

\definecolor{boxblue}{HTML}{4E79A7}
\definecolor{boxred}{HTML}{E15759}
\newcommand{\best}[1]{\textcolor{boxred}{\textbf{#1}}}
\newcommand{\second}[1]{\textcolor{boxblue}{\textbf{#1}}}

\RequirePackage{xspace}
\makeatletter
\DeclareRobustCommand\onedot{\futurelet\@let@token\@onedot}
\def\@onedot{\ifx\@let@token.\else.\null\fi\xspace}
\def\eg{\emph{e.g}\onedot} 

\def\ie{\emph{i.e}\onedot}

\makeatother

\begin{document}

\title{See Silhouettes in Motion with Neuromorphic Vision}

\author{Pei Zhang, Shijie Lin, Zhou Ge, Jinpeng Chen, and Wei Pu 
\thanks{Pei Zhang is with the School of Electrical Engineering, Guangxi University (e-mail: pei.zhang@connect.hku.hk).}
\thanks{Shijie Lin is with the Department of Computer Science, The University of Hong Kong (e-mail: lsj2048@connect.hku.hk).}
\thanks{Zhou Ge is with the School of Mechatronic Engineering and Automation, Shanghai University, and also with the SHU General Intelligent Robotics Research Institute (e-mail: gezhou@shu.edu.cn).}
\thanks{Jinpeng Chen is with the School of Computer Science (National Pilot Software Engineering School), Beijing University of Posts and Telecommunications (e-mail: jpchen@bupt.edu.cn).}
\thanks{Wei Pu is with the School of Information and Communication Engineering, University of Electronic Science and Technology of China (e-mail: puwei@uestc.edu.cn).}
\thanks{Corresponding author: Pei Zhang. Part of this work was carried out while he was at The University of Hong Kong.}}



\maketitle

\begin{abstract}
Quasi-bimodal objects, such as text, road signs, and barcodes, play a basic yet vital role in daily visual communication. By boiling these down to clear silhouettes, binarization uses a minimal language to convey essential vision cues for maximum downstream efficiency, especially for tasks that require simple geometric, topological reasoning rather than heavy appearance modeling. The catch is that frame-based imaging often struggles on mobile platforms like drones, self-driving cars, and underwater vehicles, in which rapid motion causes severe motion blur and harsh lighting washes out scene details. To overcome these physical limits, neuromorphic vision via event cameras, featuring microsecond time resolution and high dynamic range, steps in as a natural solution. Building upon this event-driven paradigm, we propose a simple yet effective dual-modal approach that harnesses the synergy between frames and events for training-free, real-time, high-frame-rate binarization on CPU-only devices. Extensive evaluations show that it earns competitive performance against leading techniques in reducing blur artifacts and delivers impressive improvements under challenging illumination at a lower computational cost. Besides, its asynchronous nature bypasses long-standing event-scarcity issues that break traditional time-binning reconstruction at fixed time slots, maintaining clear target shapes even at extreme kilohertz frame rates. Its binary results further serve as reliable representations to facilitate a range of downstream tasks. This work paves the way towards lightweight perception and interaction in embodied intelligence on resource-constrained edge platforms.
\end{abstract}


\section{Introduction}
\IEEEPARstart{C}{onsider} a common night-driving scenario in Fig.~\ref{fig:intro}: glaring high beams from a trailing car strike a reflective traffic sign in the road ahead. Just like sudden glare blinds a driver, onboard image sensors with limited dynamic range fail similarly. Due to inherent latency, the auto-exposure system cannot quickly reduce light intake. Coupled with a long shutter for nighttime low light, this delay causes partial saturation that washes out the sign while leaving motion blur and ghosting. In such urgent cases, neither classic edge detection techniques nor modern multimodal large language models (MLLMs) can give reliable feedback, as both are bottlenecked by the loss of raw visual information. Bridging this perception gap motivates our work\footnote{For review purposes, video demonstrations are attached in the supplementary materials for a better understanding of our work.}.

\input{src/intro}
We find quasi-bimodal objects everywhere in daily life, like text, road signs, and barcodes. Checkerboards and fiducials are often used for camera calibration, localization, and tracking in robotics, AR/VR, and imaging systems~\cite{wang2024cylindertag,phung2024shared,ghosh2025event}. Binarization of images or frames, which is a kind of data compression, uses a minimal language (\ie, either $0$ or $1$) to describe the essential structure of a scene for maximum downstream efficiency across inference accuracy and resource footprints, especially when tasks can be solved by simple geometric or topological reasoning rather than heavy appearance modeling. However, it is challenging to obtain clear bimodal silhouettes on mobile devices like drones, self-driving cars, and underwater vehicles, where image sensors are often blinded by complex dynamics and harsh illumination. 

To overcome the physical limits of frame-based imaging, event cameras offer a popular solution for real-time perception on mobile platforms, due to their microsecond temporal resolution, high dynamic range (HDR, over~$120$~\si{\decibel}), and low resource consumption~\cite{gallego2022event,gehrig2024low}. The sparse, binary-like nature of event streams meshes with binarization. Nevertheless, their asynchronous shape remains incompatible with most downstream algorithms, as modern hardware accelerators favor synchronous, grid-based representations~\cite{deng2021learning,zhu2026spectrogen}. Moreover, events encode intensity variations rather than absolute brightness, which is insufficient for recognizing certain objects (\eg, QR codes) where we require dense spatial context beyond mere edges. As such, frame reconstruction remains necessary. 

While one could suggest a step-wise pipeline that first goes through existing dual-modal motion deblurring methods to recover a sharp intensity frame~\cite{qi2024deblurring,yang2023event,pan2019bringing} and then uses well-established binarization techniques~\cite{mustafa2018binarization,hadjadj2016isauvola,calvo2019selectional}, this is both suboptimal and redundant. The former is often computationally intensive for its reliance on deep learning or complex optimization, and the resulting artifacts can further impair subsequent binarization, failing to meet the strict requirements for accuracy and computational efficiency. In practice, resource-constrained edge systems normally reserve their compute for demanding downstream models like MLLMs and leave only a limited budget for data pre-processing. Pre-processing should also keep pace with the rapid response of event cameras to avoid bottlenecks. This alignment preserves hardware advantages and ensures end-to-end system responsiveness. 

Given above observations, we propose a simple, lightweight, yet effective dual-modal approach that leverages the synergy between frames and events for unified, single-step, training-free binarization. Through self-adaptive modulation, it delivers faithful binary results under severe motion blur and harsh illumination. With linear time complexity, it comfortably achieves real-time performance on CPU-only platforms. Besides, its asynchronous nature bypasses long-standing event-scarcity issues, which break traditional time-binning reconstruction at fixed time slots, and decouples computational overhead from frame rates. As such, it can preserve clear target shapes with consistently low latency even at extreme kilohertz scales. The resulting binary representations also benefit downstream tasks like text recognition, object tracking, and optical flow estimation. Such a solution holds strong potential to bridge the gap between high-speed sensing and efficient edge computation.

The remainder of this paper is organized as follows. Section~\ref{sec:work} reviews related work. Section~\ref{sec:method} presents the proposed dual-modal binarization method. Section~\ref{sec:ex} describes the experimental design and reports comparative evaluations against existing techniques. Section~\ref{sec:end} concludes the work with its limitations and future endeavors.

\section{Related Work}\label{sec:work}
\subsection{Image Binarization}
Image binarization, which can be taken as data compression, is a fundamental pre-processing for high-level vision tasks. Early approaches hinge on global statistical analysis~\cite{otsu1975threshold,mustafa2018binarization} or local adaptive thresholding~\cite{su2012robust,khurshid2009comparison} to separate foreground from background. While computationally efficient, they often struggle with texture degradation under challenging illumination and then leave unstable performance. The advance of deep learning shifts the focus towards semantic segmentation, where convolutional neural networks (CNNs) learn hierarchical representations across either pixels or patches~\cite{he2019deepotsu,calvo2019selectional}. Despite their improved robustness, these frame-based methods remain constrained by the limited temporal resolution of traditional image sensors. Thus, they often bring structural distortions, loss of fine details, and temporal inconsistency among frames, when motion blur is treated as spatial artifacts to be painfully reconstructed instead of temporal signals to be naturally captured~\cite{qu2025self}. Such challenges make frame-based solutions impractical for real-world dynamic scenes beyond controlled lab conditions~\cite{sulaiman2019degraded}.

\subsection{Event Cameras}
Event cameras pose a paradigm shift in visual perception. As opposed to frame-based imaging that captures absolute intensity at a fixed frame rate, they asynchronously measure per-pixel changes in brightness and generate continuous event streams in response to scene dynamics~\cite{gallego2022event}. Naturally, the resulting sparse, spike-like streams are incompatible with mainstream vision frameworks that favor synchronous representations, which has been driving the search for alternative modeling techniques~\cite{duan2025eventaid}. Such a biologically inspired mechanism offers several attractive features, including microsecond-level temporal resolution, HDR, ultra-low latency, and minimal resource consumption. Accordingly, they have received growing attention across a wide range of domains, including agile aerial robotics~\cite{falanga2020dynamic,magrini2025drone}, high-speed advanced manufacturing~\cite{zhu2025ultrafast,ge2023millisecond,wang2025angle}, quantum sensing~\cite{du2024widefield}, and biological research~\cite{guo2024eventlfm,zhang2025tcsvt,chen2026self}. With the rapid rise of embodied intelligence driving the need for real-time, energy-efficient visual systems, event cameras provide an alternative for robust performance under challenging imaging conditions~\cite{lin2024embodied,wang2026event}.

\input{src/intro2}
\subsection{Dual-Modal Motion Deblurring}
Event cameras offer a viable solution to bypass the temporal limitations of frame-based sensors, enabling a blur-free capture of high-speed motion under challenging illumination. To bridge such a precise measurement with visual reconstruction, various computational paradigms have been developed. Model-based approaches, which establish physical mappings between event streams and latent sharp frames, typically offer fast responsiveness, CPU-only feasibility, and rigorous interpretability~\cite{scheerlinck2018continuous,pan2019bringing, zhang2024tip}. Concurrently, the rapid advance of learning-based techniques fosters the exploration of dual-modal frame-event synergy for motion deblurring, spanning from discriminative backbones (\eg, CNNs~\cite{rebecq2021high,sun2022event,wu2026dark}, Transformers~\cite{xu2025motion}, Mamba~\cite{xiao2025event}) to emerging generative strategies (\eg, diffusion~\cite{xie2025diffusion,yang2025event}). While these methods earn state-of-the-art reconstruction fidelity, they often come at the cost of prohibitive computational complexity and introduce substantial redundancy for binary representations. Consequently, they are ill-suited for efficient binarization and fail to meet the strict speed and resource requirements for lightweight applications.

\subsection{Dual-Modal Binarization}
Directly deriving binary motion signals from frame-event synergy is a shift to bypass the high computational overhead of full-scale intensity reconstruction. Despite its potential, research in this branch remains nascent. One pioneering effort frames it as pixel-wise binary classification yet suffers from severe performance degradation under non-ideal volatile illumination\cite{lin2024neuromorphic}. Such sensitivity to lighting conditions restricts its deployment in unpredictable real-world environments. Beyond that, the industrial significance of this task is recently highlighted by the integration of event cameras into smart glasses for low-power optical character recognition (OCR)~\cite{brander2025reading}. Thus, developing a fast, lightweight, dual-modal binarization framework resilient across diverse lighting is not an incremental improvement, but a prerequisite for reliable real-time vision on resource-constrained edge platforms.

\section{Methodology}\label{sec:method}
\input{src/dual}
\subsection{Dual-Modal Binarization Framework}
\subsubsection{Problem Definition}
Our objective is to recover a temporally dense, sharp binary video sequence from motion-blurred intensity frames and their corresponding asynchronous events. The proposed workflow shown in Fig.~\ref{fig:intro2} follows coherent three steps for final high-frame-rate results, including a spatial decomposition framework, adaptive parameter modulation, and asynchronous state propagation. 

We define the latent sharp scene radiance as $L(\mathbf{x}, t)$, where $\mathbf{x}$, $t$ denote spatial coordinates and time, respectively. The active-pixel sensor (APS) accumulates photons over an exposure interval $\mathcal{T} = [t, t+\tau]$ with duration $\tau$. In a dynamic scene, the captured intensity frame $I(\mathbf{x})$ represents a temporal integration of the time-varying scene radiance, often exhibiting severe motion blur
\begin{equation}\label{eq:I}
    I(\mathbf{x}) = \frac{1}{\tau} \int_{\mathcal{T}} L(\mathbf{x}, t)~\mathrm{d}t.
\end{equation}
For a bimodal scene, $L(\mathbf{x}, t)$ ideally comprises two intensity levels, foreground and background. However, temporal integration along motion trajectories destroys high-frequency spatial cues essential for accurate binarization.

In parallel, the dynamic vision sensor (DVS)\footnote{For example, the DAVIS346 event camera integrates a DVS and an APS within the same pixel array, with each pixel sharing a single photodiode~\cite{taverni2018front}.} generates a stream of events $\mathcal{E} = \{e_k\}_{k=1}^{K}$, where $K$ is the cardinality of $\mathcal{E}$. Each event $e_k = (\mathbf{x}_k, t_k, p_k)$, indexed by $k$, is triggered asynchronously when the logarithmic intensity at $\mathbf{x}_k$ changes by a temporal contrast $c$
\begin{equation}\label{eq:E}
    \ln L(\mathbf{x}_k, t_k) - \ln L(\mathbf{x}_k, t_k - \Delta t_k) = p_k c,
\end{equation}
where $\Delta t_k$ is the time elapsed since the last event at $\mathbf{x}_k$, and $p_k \in \{+1, -1\}$ is the polarity. In this case, dual-modal binarization seeks a mapping $f$ that reconstructs a sharp binary frame $B(\mathbf{x}, t) \in \{0, 1\}$ for any time $t \in \mathcal{T}$ from both the blurry observation and event stream  
\begin{equation}\label{eq:B}
    B(\mathbf{x}, t) = f \big(I(\mathbf{x}), \mathcal{E}; t \big).
\end{equation}

Motion blur corrupts the intensity integration at transitional boundaries, whereas static regions remain consistent. To formulate binarization as activity-based per-pixel classification, we decompose the spatial domain $\mathrm{\Omega} = \mathbb{D} \cup \mathbb{S}$ into disjoint dynamic and static subsets, $\mathbb{D}$ and $\mathbb{S}$, as shown in Fig.~\ref{fig:dual}. Pixels $\mathbf{x} \in \mathbb{D}$ correspond to locations experiencing a significant radiance transition during $\mathcal{T}$, where $I(\mathbf{x})$ becomes a mixture of foreground and background values. Conversely, for $\mathbf{x} \in \mathbb{S}$, the integration in Eq.~\eqref{eq:I} degenerates to a scaling of the constant latent state, making $I(\mathbf{x})$ a valid measure for binarization. We then reformulate $f$ as the union of two independent inference operators, $f_{\mathcal{E}}$ (event-driven) and $f_{I}$ (intensity-based), targeting a specific reference time $t' \in \mathcal{T}$
\begin{equation}
\begin{aligned}
B(\mathbf{x}, t') &= f \big(I(\mathbf{x}), \mathcal{E}; t' \big)\\
&= \mathbb{I}_{\mathbf{x} \in \mathbb{D}}\,f_{\mathcal{E}}(\mathcal{E}; \mathbf{x}, t')
+ \mathbb{I}_{\mathbf{x} \in \mathbb{S}}\,f_I\big(I(\mathbf{x})\big).
\end{aligned}
\end{equation}
Here, $f_{\mathcal{E}}$ leverages the microsecond-level temporal resolution of events to decode a binary state from moving edges aligned to $t'$, while $f_{I}$ resolves static regions based on global intensity statistics. Then, this formulation isolates an ill-posed motion deblurring problem to the sparse $\mathbb{D}$ naturally captured by event cameras and thus improves computational efficiency.

\subsubsection{Event-Driven Inference} We recover $B(\mathbf{x}, t')$ for $\mathbf{x} \in \mathbb{D}$ by integrating events to detect significant polarity shifts. Let $I_\mathcal{E}(\mathbf{x}, t')$ be the accumulated logarithmic intensity change
\begin{equation}\label{eq:log}
I_\mathcal{E}(\mathbf{x},t') = c \sum_{e_k \in \mathcal{E}} p_k \delta(\mathbf{x} - \mathbf{x}_k)\mathcal{H}(t'-t_k),
\end{equation}
where $\delta$ is the Kronecker delta, and $\mathcal{H}$ denotes the Heaviside step function. By causality, a brightness rise implies a pixel started from dark ($0$), and a drop implies it started from bright ($1$). We then define $f_{\mathcal{E}}$ by an event-driven confidence $\theta_\mathcal{E}$
\begin{equation}
f_{\mathcal{E}}(\mathcal{E}; \mathbf{x}, t') =
\begin{cases}
0 & \text{if } I_\mathcal{E}(\mathbf{x},t') \ge \theta_\mathcal{E}, \\
1 & \text{if } I_\mathcal{E}(\mathbf{x},t') \le -\theta_\mathcal{E}.
\end{cases}
\end{equation}
Pixels (\eg, where noise exists) that do not accumulate sufficient contrast (\ie, $|I_\mathcal{E}| < \theta_\mathcal{E}$) despite being in $\mathbb{D}$ are treated as ambiguous and then reassigned to the intensity-based inference branch. 

\subsubsection{Intensity-Based Inference} For $\mathbf{x} \in \mathbb{S}$, $I(\mathbf{x})$ remains a reliable measure for the latent state. We apply an intensity binarization threshold $\theta_I$
\begin{equation}
f_{I}\big(I(\mathbf{x})\big) =
\begin{cases}
1 & \text{if } I(\mathbf{x}) > \theta_I, \\
0 & \text{otherwise}.
\end{cases}
\end{equation}
This dual-modal binarization framework leaves three parameters $c$, $\theta_{\mathcal{E}}$, $\theta_I$ up in the air. We discuss their estimation in the context of binarization in the following sections.

\input{src/contrast}
\subsection{Adaptive Parameter Estimation}
\subsubsection{Online Temporal Contrast Estimation}\label{sec:c}
Assuming $c$ to be a time-invariant constant fails in real-world scenes where DVS sensitivity drifts frequently, which often leads to reconstruction bias~\cite{zhang2024tip}. Observing that global shifts dominate stochastic inter-pixel mismatches, we bypass costly per-pixel calibration in favor of adaptive sensitivity tracking to rapidly stabilize event computation under lighting fluctuations. 

As such, we propose an online log-domain variance matching estimator. Our key insight builds on the statistical stability of quasi-bimodal scenes, in which the photometric spread is dominated by inter-class separation between foreground and background rather than intra-class texture variations. While motion blur smooths high-frequency edges, the global statistical dispersion, anchored by the dominant intensity peaks, remains sufficiently informative for contrast estimation as long as histogram separability is not fully destroyed. As shown in Fig.~\ref{fig:contrast}, motion blur does not reshape the distribution of both frames and events yet reduces their variances~\cite{gallego2018unifying,zhang2024event}.
Although they are not pixel-wise equivalent, in quasi-bimodal dynamic regions, their spatial dispersions are driven by the same foreground-background transitions. We thus relate them statistically by matching their second-order moments over $\mathbb{D}$ to estimate a contrast
\begin{equation}\label{eq:contrast}
    \sigma^2_{\mathbb{D}}\Big(\ln\big(G_{\tilde{\sigma}} * I(\mathbf{x}) + \epsilon\big)\Big) \approx c^2 \cdot \sigma^2_{\mathbb{D}}\big(E(\mathbf{x}, t')\big), 
\end{equation}
viewed as a lightweight energy-matching problem
\begin{equation}
    \hat{c} = \arg\min_{c>0}\bigg[
    \sigma^2_{\mathbb{D}}\Big(\ln\big(G_{\tilde{\sigma}} * I(\mathbf{x}) + \epsilon\big)\Big) - c^2\sigma^2_{\mathbb{D}}\big(E(\mathbf{x},t')\big)\bigg]^2,
\end{equation}
where $G_{\tilde{\sigma}}$ denotes a Gaussian kernel with a scale $\tilde{\sigma}$ to suppress sensor artifacts, $\epsilon$ is a small offset, and $E(\mathbf{x},t') \triangleq \sum p_k\delta(\mathbf{x}-\mathbf{x}_k)\mathcal{H}(t'-t_k)$. We also restrict the operation to $\mathbb{D}$ to improve computational efficiency and avoid instability in static regions. This scale alignment gives a closed-form estimate
\begin{equation}
     \hat{c} = \Big[\sigma^2_{\mathbb{D}}\!\Big(\ln\big(G_{\tilde{\sigma}} * I(\mathbf{x}) + \epsilon\big)\Big)/\sigma^2_{\mathbb{D}}\!\big(E(\mathbf{x},t')\big)\Big]^{1/2}.
\end{equation}
Per-frame estimation is prone to cause temporal flickering. We then update $c$ by an exponential moving average across frames
\begin{equation} \label{eq:ema}
    c_n = \alpha \cdot \hat{c}_n + (1 - \alpha) \cdot c_{n-1},
\end{equation}
where $n$ indexes frames, and $\alpha$ controls the adaptation rate. This temporal smoothing prevents abrupt scale drift and preserves inter-frame consistency when instantaneous estimates become unreliable under extreme illumination. Unlike iterative optimization~\cite{pan2019bringing}, our linear-time closed-form solution offers real-time adaptability to illumination changes without offline calibration or pre-training~\cite{wang19event}.

\input{src/lambda}
\subsubsection{Dynamic Range Reshaping}\label{sec:fusion}
High-quality binarization hinges on histogram separability. However, $I(\mathbf{x})$ may undergo saturation or photon noise for overexposure or underexposure, causing the histogram to be flat or unimodal where structures are buried. To recover separability, we propose an exposure-adaptive fusion strategy for dynamic range reshaping.

We quantify intensity reliability $\lambda = \exp\big(-\beta (\mu_I - \mu)^2\big) \in (0, 1]$ based on the deviation of the mean $\mu_I$ of $I(\mathbf{x})$ from an ideal mid-gray $\mu$, where $\beta $ controls sensitivity. A $\lambda \to 1$ indicates well-balanced exposure, while $\lambda \to 0$ signals distribution collapse. In this sense, $\lambda$ acts as a confidence weight for the degraded $I(\mathbf{x})$. Let $J$ denote a candidate reshaped frame. We obtain $I_\lambda$ as the closed-form result of balancing the observed intensity and an event-driven prior
\begin{equation}\label{eq:fusion_min}
     I_\lambda = \arg\min_J \lambda \|J-I\|_2^2 +(1-\lambda)\big\|J-(\mathbb{I}_{\mathbf{x} \in \mathbb{D}}\cdot \tilde{I}_\mathcal{E}+\mathbb{I}_{\mathbf{x} \in \mathbb{S}}\cdot \mu)\big\|_2^2,
\end{equation}
which yields a simple convex fusion
\begin{equation}\label{eq:fusion}
     I_\lambda(\mathbf{x},t') = \lambda I(\mathbf{x}) + (1-\lambda)\big(\mathbb{I}_{\mathbf{x} \in \mathbb{D}}\cdot \tilde{I}_\mathcal{E}(\mathbf{x},t') + \mathbb{I}_{\mathbf{x} \in \mathbb{S}} \cdot \mu\big),
\end{equation}
where $\tilde{I}_\mathcal{E} = \exp(I_\mathcal{E})$. This reshapes histogram separability via two complementary effects. In dynamic areas where motion blur and exposure imbalance co-occur, the HDR event prior restores degraded structural features. As $\lambda \to 0$, the fusion prioritizes events to recover missing contrast. Meanwhile, in static regions where poor exposure manifests as high-frequency photon noise or clamping, pixels are softly pulled towards a neutral pivot to suppress unreliable variations caused by exposure imbalance. As a result, $I_\lambda$ enhances histogram separability by reinforcing inter-class contrast and suppressing intra-class dispersion, which facilitates the identification of an optimal separating threshold even when intensity frames are severely degraded. Fig.~\ref{fig:lambda} illustrates an OCR scene~\cite{zhang2024tip} and its corresponding distribution to explain the process.

\textbf{Observability and Extreme Cases.} Eq.~\eqref{eq:contrast} assumes that $I(\mathbf{x})$ retains sufficient spatial observability. Both log-domain compression and temporal smoothing retain reliable contrast scaling under moderate exposure imbalance, but severe saturation is prone to bias variance estimates. The issue becomes ill-posed only when both intensity frames and events lose perceptual information (\eg, bright glare with slight motion, as letters in Fig.~\ref{fig:limit}), leaving no recoverable contrast cues in either modality. Notably, Eqs.~\eqref{eq:contrast} and~\eqref{eq:fusion_min} operate at distinct layers more than a dependency. Estimating $c$ calibrates event amplitudes to the intensity scale across varying illumination, which maximizes histogram separability for precise binarization. Even if $c$ degrades for extreme exposure, the bias affects contrast magnitude rather than topological integrity. As a binarization task targets inter-class separability instead of exact radiance, moderate scale deviations do not induce misclassification. Failure occurs only at the observability limit, where neither modality conveys sufficient discriminative information.

\input{src/limit}
\subsubsection{Optimal Threshold Estimation}
Once reshaped, $I_\lambda(\mathbf{x},t')$ yields a more separable bimodal distribution for thresholding. Traditional Otsu estimates a threshold $\theta$ by maximizing the inter-class variance $\sigma_b^2$~\cite{otsu1975threshold}. However, it is histogram-based and spatially blind. When residual noise, saturation artifacts, or event-induced fluctuations shift histogram peaks, a pure statistical threshold may deviate from physical object boundaries, leading to suboptimal results.

To incorporate the restored structures in $I_\lambda(\mathbf{x},t')$ without extra optimization, we formulate threshold selection from a lightweight MAP view. Specifically, $\sigma_b^2(\theta)$ statistically quantifies histogram separability, while the gradient density around each candidate threshold characterizes boundary consistency. For each candidate $\theta$ over quantized intensity bins, we define
\begin{equation}
   g(\theta) = \sum_{\mathbf{x}: I_\lambda(\mathbf{x},t') = \theta} \|\nabla I_\lambda(\mathbf{x},t')\|
\end{equation}
to measure the accumulated edge strength of pixels falling into $\theta$. A large $g(\theta)$ indicates that the corresponding intensity level is close to strong structural transitions, and is therefore more likely to lie near a true foreground-background boundary. Then, we seek a threshold favoring both statistical separability and structural consistency. Under this MAP-inspired view, the posterior score can be written as
\begin{equation}
    \mathbb{P}(\theta|I_\lambda) \propto \mathbb{P}_{\mathrm{stat}}(\theta|I_\lambda) \mathbb{P}_{\mathrm{grad}}(\theta|I_\lambda),
\end{equation}
where $\mathbb{P}_{\mathrm{stat}}$ and $\mathbb{P}_{\mathrm{grad}}$ are the scores from histogram statistics and edge structures, respectively. Taking the logarithm gives
\begin{equation}
\begin{aligned}
\theta^* &= \arg\max_\theta \ln \mathbb{P}(\theta|I_\lambda) \\
&= \arg\max_\theta \ln \mathbb{P}_{\mathrm{stat}}(\theta|I_\lambda) + \ln \mathbb{P}_{\mathrm{grad}}(\theta|I_\lambda).
\end{aligned}
\end{equation}
We instantiate these two terms with $\mathbb{P}_{\mathrm{stat}}(\theta|I_\lambda) \propto \sigma_b^2(\theta)+\epsilon_b$
and $\mathbb{P}_{\mathrm{grad}}(\theta|I_\lambda) \propto 1+g(\theta)$, which leads to
\begin{equation}\label{eq:thes_est}
\theta^{*}=\arg\max_{\theta}\ln\big(\sigma_b^2(\theta)+\epsilon_b\big)+\ln\big(1+g(\theta)\big),
\end{equation}
where $\epsilon_b$ is a small offset. This form preserves the efficiency of Otsu thresholding while adding a structural prior that favors thresholds aligned with restored edges.

While $\theta^*$ provides a global baseline, slight deviations may still arise due to the quantization of intensity levels. We thus refine $\theta_I$ by searching for the strongest gradient response within a narrow $2\%$ neighborhood $\mathcal{N}(\theta^*)$
\begin{equation}
\theta_I = \arg\max_{\theta \in \mathcal{N}(\theta^*)} g(\theta).
\end{equation}
Snapping $\theta_I$ to the nearest dominant gradient ridge improves boundary alignment without sacrificing the efficiency of global thresholding. To maintain dual-modal consistency, we calibrate $\theta_{\mathcal{E}}$ by mapping $\theta_I$ to the event domain
\begin{equation}
    \theta_{\mathcal{E}} = \theta_I \cdot \max_{\mathbf{x} \in \mathbb{D}} |I_\mathcal{E}(\mathbf{x}, t')|.
\end{equation}
Here, $\theta_I$ specifies a normalized separating position, and this mapping rescales it to the event domain such that $f_{\mathcal{E}}$ and $f_I$ operate on a unified decision space. This statistical-structural coupling resolves the separating ambiguity to ensure consistent binarization performance across varying imaging conditions.

\input{src/algorithm}
\subsection{High-Frame-Rate Video Generation}
With $c$, $\theta_I$, $\theta_\mathcal{E}$ determined, we obtain a static binary snapshot $B(\mathbf{x}, t)$ at any time. However, recovering the continuous temporal evolution of a scene via frame-by-frame reconstruction is computationally redundant and prone to temporal jitter. As such, we propose an asynchronous state propagation strategy that treats the initial binary frame as a seed and propagates it asynchronously to the temporal domain, driven solely by inter-frame events. This reformulates a video generation problem from dense reconstruction into a sparse, event-driven update process, thus minimizing latency and computational load.

\input{src/ex_binary_normal}
\input{src/ex_binary_low}
As shown in Algorithm~\ref{alg:asp}, we model the binary transition of each pixel as a state-dependent accumulation process\footnote{We set $\rho_k = (1+p_k)/2$ to map $p_k \in \{+1,-1\}$ to $\rho_k \in \{1,0\}$.}. Instead of complex optimization, we only track a residual $r(\mathbf{x}, t)$ that records incremental changes relative to the current state. For each incoming event, we update the residual at a specific $\mathbf{x}$ based on its prior state
\begin{equation} 
    r(\mathbf{x}, t_k) = r(\mathbf{x}, t_{k-1}) + p_k \cdot c \cdot \mathbb{I}[B(\mathbf{x}, t_{k-1}) \oplus \rho_k], 
\end{equation}
where the indicator function activates only when $p_k$ opposes the current state. If a pixel is already bright, we ignore positive events and only accumulate negative ones. It also functions as noise filtering where small radiance fluctuations or redundant events too few to trigger a flip are suppressed. Then, $B(\mathbf{x}, t_k)$ remains unchanged until the residual sufficiently accumulates 
\begin{equation} 
B(\mathbf{x}, t_k) = 
\begin{cases} 1 - B(\mathbf{x}, t_{k-1}) & \text{if}~|r(\mathbf{x}, t_k)| \ge \theta_{\mathcal{E}}, \\ B(\mathbf{x}, t_{k-1}) & \text{otherwise}. 
\end{cases} 
\end{equation}
$r(\mathbf{x}, t_k) $ is reset upon a state flip to ensure each flip corresponds to a separate binary transition and to prevent one strong edge from causing repeated false flips. To suppress salt-and-pepper artifacts, each flip triggers a local consistency check within a $3 \times 3$ window $\mathcal{W}(\mathbf{x})$ centered at the flipped pixel. Its state is updated by majority voting
\begin{equation}
    B(\mathbf{x}, t_k)  \leftarrow  \mathbb{I} \Bigg[\frac{\sum_{\mathbf{x}' \in \mathcal{W}(\mathbf{x})} B(\mathbf{x}', t_k)}{|\mathcal{W}(\mathbf{x})|} > \frac{1}{2} \Bigg],
\end{equation}
which suppresses isolated flips that disagree with their local spatial context while preserving changes supported by neighboring pixels. Since it performs asynchronously and sparsely only at the location of each flip, it avoids full-frame scanning and requires only constant computation per update.

In a nutshell, we keep a continuously evolving state $B(\mathbf{x}, t)$ in memory at the microsecond temporal resolution of events. Then, high-frame-rate video sequences can be generated by simply sampling this state at any desired frequency without recomputation. In this view, the video is essentially a sequence of snapshots of this evolving state, by which we decouple reconstruction frame rates from computational load.

\subsection{Time Complexity Analysis}
We analyze computational complexity to validate the real-time capability of our method. Let $N$ denote spatial resolution. First, the dual-modal binarization framework incurs a cost of $\mathcal{O}(K+N)$ for event integration and pixel-wise classification over the whole spatial domain. Similarly, the log-domain variance matching estimator bypasses iterative solvers and relies on closed-form statistical moments over the sparse subset of pixels $\mathbb{D}$, thus limiting its complexity to a linear $\mathcal{O}(K+N)$. Both dynamic range reshaping and the gradient-aware Otsu estimator function as lightweight statistical passes, with each contributing bounded $\mathcal{O}(N)$ complexity.

Unlike conventional synchronous methods that scale with $\mathcal{O}(N \times \text{FPS)}$ complexity\footnote{FPS = frames per second.}, our asynchronous video generation, driven by sparse events, scales purely with the scene dynamics $\mathcal{O}(K)$. For an event camera (\eg, DAVIS346, $N = 346\times260$), $K$ represents sparse temporal changes rather than redundant dense frames. As such, our algorithm decouples computational cost from output frame rates and allows for the generation of a kilohertz-equivalent video with consistently low latency, fitting within high-speed real-time pipelines.

\input{src/ex_binary_glare}
\input{src/ex_reconstruction}
\section{Experiment}\label{sec:ex}
We use automatically calibrated $c$, $\theta_I$, $\theta_\mathcal{E}$ and empirically set $\alpha=0.2$, $\beta=20$ for binarization. Public datasets CF~\cite{scheerlinck2018continuous}, HQF~\cite{stoffregen2020reducing}, REBlur~\cite{sun2022event}, EBT~\cite{lin2024neuromorphic}, and RND~\cite{zhang2024tip} are used for qualitative or quantitative comparisons. To avoid ambiguity in ground truth, evaluations are restricted to the samples that exhibit a clear bimodal distribution. Following standard practice~\cite{lin2024neuromorphic}, the ground truth for quantitative comparisons is generated from available sharp frames, where we estimate a motion-invariant threshold and refine it for accurate bimodal separation~\cite{otsu1975threshold}. Numerically, we report three widely used metrics, including the Matthews Correlation Coefficient (MCC) that measures balanced binary classification performance and remains reliable under class imbalance~\cite{chicco2020advantages} 
\begin{equation}
    \text{MCC} = \frac{\text{TP} \cdot \text{TN} - \text{FP} \cdot \text{FN}}{\sqrt{(\text{TP}+\text{FP})(\text{TP}+\text{FN})(\text{TN}+\text{FP})(\text{TN}+\text{FN})}},
\end{equation}
where the above abbreviations are four confusion matrix terms, the Peak Signal-to-Noise Ratio (PSNR) that quantifies pixel-wise consistency with the ground truth
\begin{equation}
    \text{PSNR} = 10 \log_{10} \left(\frac{\text{MAX}_I^2}{\text{MSE}}\right),
\end{equation}
where $\text{MAX}_I^2 = 1$ and $\text{MSE} = (\text{FP}+\text{FN})/(\text{TP}+\text{TN}+\text{FP}+\text{FN})$ in binarization tasks, and the Negative Rate Metric (NRM) that captures pixel-wise misclassification rates
\begin{equation}
    \text{NRM} = \frac{1}{2}\left(\frac{\text{FN}}{\text{TP}+\text{FN}} + \frac{\text{FP}}{\text{FP}+\text{TN}}\right).
\end{equation}
The measured values are averaged over the selected samples. For those with well-balanced exposure, We further augment real-world motion blur with nighttime low light (with sensor noise) and bright glare from close-range active illumination, as various challenges for robustness evaluations.

\input{src/quan_comparison}
\subsection{Binarization Results}
Figs.~\ref{fig:ex_binary_normal},~\ref{fig:ex_binary_low}, and~\ref{fig:ex_binary_glare} visually compare ours with state-of-the-art techniques, including frame-only model-based~\cite{khurshid2009comparison,hadjadj2016isauvola,mustafa2018binarization}, learning-based~\cite{calvo2019selectional} methods, and a dual-modal one~\cite{lin2024neuromorphic}. Our evaluation covers three challenging scenes: severe motion blur, motion blur with low light, and with bright glare. Frame-only ones, like human vision, degrade markedly across almost all cases once the frames lose informative visual cues as a result of spatial smoothing or intensity clipping. While the dual-modal competitor alleviates motion blur, it remains quite susceptible to harsh lighting. Comparatively, our method consistently recovers sharp and faithful object boundaries across all test cases even when motion and photometric conditions are adverse, proving strong robustness to mixed degradations. Specifically, although the checkerboard (Fig.~\ref{fig:ex_binary_normal}) exhibits minor blur, they fail to delineate the four corners of black squares as clearly as ours. In low lighting (Fig.~\ref{fig:ex_binary_low}, 1st sample), both the frame and events are filled with noise that is amplified by the counterparts but suppressed in ours. These results highlight the strength of our method in maintaining reliable performance under challenging real-world environments.

Fig.~\ref{fig:ex_reconstruction} extends the match by comparing ours with state-of-the-art motion deblurring methods, including frame-only~\cite{bai2018graph,chen2019blind} and dual-modal ones~\cite{scheerlinck2018continuous,pan2019bringing,zhang2024tip}. It features two representative scenarios: non-uniform motion blur (top) and a real-world pedestrian scene under low-light conditions with partial glare (bottom). For fair, all deblurred results are binarized using the same established baseline~\cite{mustafa2018binarization}\footnote{We use this setting for all step-wise pipelines.}. As illustrated, frame-only competitors struggle to have an informative reconstruction in the absence of motion and HDR cues to reveal which pixels are changing or hidden behind. While dual-modal deblurring methods improve via complementary event features, their resulting intensity frames still exhibit residual artifacts that propagate as errors and compromise the quality of subsequent binarization. This reveals the limitation of step-wise pipelines where errors accumulate across stages. In contrast, our unified framework bypasses intermediate reconstruction efforts and error propagation to directly estimate a binary representation. It thereby consistently delivers more faithful binarization and shows reliability in real-world scenarios with complex motion patterns and mixed degradations.

\input{src/ex_downstream}
\input{src/quan_downstream}

Table~\ref{tab:quan_comparison} compares quantitative results across three datasets under varying illumination. Consistent with the visual observations, dual-modal methods outperform frame-only ones by a clear margin. Under moderate lighting, our method achieves near the state-of-the-art and remains highly competitive with the strong baseline~\cite{lin2024neuromorphic}. More importantly, it delivers impressive results under harsh imaging conditions. In challenging low light and bright glare scenarios where the performance of the others drops sharply, it consistently leads the runner-up across all evaluation metrics by a wide margin.

\input{src/ex_framerate}
\subsection{Downstream Tasks}
High-quality binarization provides a compact yet informative source that benefits a range of downstream tasks in both accuracy and computational efficiency. Fig.~\ref{fig:ex_downstream} shows the superiority of our approach across three representative applications, including OCR, fiducial marker tracking, and optical flow estimation, under diverse illumination. These tasks often place strict demands on input quality, where minor distortions or information loss can clearly compromise performance. We exploit the popular learning-based techniques EasyOCR~\cite{baek2019character,shi2016end}, DeepTag~\cite{zhang2022deeptag}, and RAFT~\cite{teed2020raft} for downstream analysis. As shown, the degraded frames and counterparts produce low-quality, ambiguous inputs to mislead downstream algorithms, leading to recognition faults in OCR, imprecise corner localization in marker tracking, and unfaithful, inconsistent boundary preservation in optical flow estimation. Comparatively, our method yields clean and structurally faithful binary representations that preserve critical scene geometry, enabling reliable text recognition, stable keypoint localization for fiducials, and coherent boundary-aware motion estimation. 

Table~\ref{tab:quan_downstream} further reports quantitative numbers for these downstream tasks, averaged over six frames captured at different time. We use the Character Error Rate (CER) to evaluate the fraction of incorrect letters, the Root Mean Square Error (RMSE) to quantify pixel-wise localization errors of detected marker corners, and the Average Endpoint Error (AEPE) to measure the deviation of predicted optical flow fields from ground truth. In line with the visual observations, frame-only methods struggle across all tasks even under gentle lighting due to unresolved motion blur, whereas dual-modal counterparts improve by a clear margin. Specifically, our approach yields sharp, reliable binary representations that keep lower error rates across three applications. These findings suggest that our solution has strong potential to serve as a robust and efficient foundation for a wide range of downstream visual tasks, especially under challenging imaging conditions.

\input{src/ex_runtime}
\subsection{High-Frame-Rate Video Analysis}
Fig.~\ref{fig:ex_framerate} shows a stress test on how frame rates influence binarization quality and downstream OCR performance, where our visual results at three timestamps are given along with the comparisons against representative EDI~\cite{pan2019bringing} on the MCC and CER metrics, evaluated over six frames at different time. These curves reveal a clear advantage of our asynchronous event-driven update over the synchronous time-binning. As frame rates grow, EDI degrades significantly since a narrowing time window reduces event counts available for computation, leaving ineffective temporal integration and poor reconstruction. In comparison, our asynchronous operation maintains a continuously evolving state, from which output frames are obtained as snapshots, to bypass the information scarcity of fixed temporal slices and then enable consistent binary results even at kilohertz frame rates. This proves that our scalable approach better aligns with the asynchronous nature of event cameras. Moreover, a slight dip is observed at $5000$ FPS. Rather than an algorithmic flaw, it reflects sensor-level constraints, including timestamp jitter and readout bandwidth at high event rates, as well as the effect of capturing asynchronous update mid-transition. These results confirm that our method scales gracefully with extreme frame rates while preserving stable binarization and downstream task reliability.

A clear distinction should be made between event scarcity at a scene level and within a fixed time slot. Naturally, dual-modal methods struggle when a scene triggers very few events (\eg, a near-static  scene), since they basically revert to frame-only baselines. This study examines the contribution of events per time slot in scenarios already filled with abundant event activity, where we focus on the ability of efficient information aggregation under limited time resolution. While our method can theoretically scale to much higher FPS (bounded by event cameras), we report our findings at $5000$ FPS, which already satisfies most real-world applications.

\subsection{Runtime Analysis}
Fig.~\ref{fig:ex_runtime} presents a runtime evaluation across four datasets, each comprising six sequences with varying temporal lengths and event counts (\ie, event rates for sparse or dense scenes). All evaluated methods run on an Apple M4 Pro chip using a single CPU core without GPU acceleration, to ensure fair hardware-consistent comparisons. We report four metrics: the total runtime, real-time factor (sequence length$/$runtime), processing rate (event counts$/$runtime), and processing latency (runtime$/$event counts). A real-time factor higher than $1$ indicates that a method processes an event stream faster than its physical acquisition time and thus operates in real time. 

Our approach minimizes runtime to perform on par with the fastest baseline~\cite{lin2024neuromorphic} and keeps real-time performance across all tested cases with either sparse or dense event streams, while the two step-wise pipelines take more runtime. Our method also sustains a high event processing throughput, making it possible to operate in step with event cameras. The relative efficiency gains observed in this controlled setting indicate our clear advantage for real-time deployment on edge devices.

\input{src/ex_ablation}
\input{src/quan_ablation}
\subsection{Ablation Study}
We conduct an ablation study to analyze the contribution of the key components in our dual-modal binarization framework. Specifically, we investigate: (i) manually setting $c$ to values lower ($\downarrow$) or higher ($\uparrow$) than the adaptively estimated one, (ii) varying the strength of dynamic range reshaping, and (iii) disabling gradient priors during threshold estimation. As Fig.~\ref{fig:ex_ablation} shows, setting a higher $c$ amplifies noise and a lower value weakens structure reshaping. Similarly, suboptimal fusion between degraded frames and events introduces distinct artifacts. Insufficient event contribution results in structural loss due to highlight clipping, while excessive reliance on sparse events causes spatial jitter and fragmented, discontinuous contours. These findings highlight the value of adaptive parameter estimation over a fixed setting across diverse scenarios. Besides, incorporating the proposed statistical–structural coupling during threshold estimation ensures more faithful binarization, particularly under challenging photometric conditions. Table~\ref{tab:quan_ablation} quantitative results are consistent with the visual observations and provide clear evidence of the effectiveness of each component in the overall framework.

\input{src/ex_quality}
\subsection{Event Quality for Binarization}
Dual-modal fusion is influenced not only by the quality of degraded frames but also by event streams. However, sensitive event cameras are susceptible to environmental interference and then respond to complex scene dynamics with non-ideal events~\cite{zhang2023neuro}. This motivates a research on how event quality impacts binarization performance. In Fig.~\ref{fig:ex_quality}~(a), we present results based on the events with real-world background activity noise from flickering lights. Such false events dominate low-frequency regions and corrupt the dual-modal fusion process. As a result, the sharp intensity frame of Zhang~\cite{zhang2024tip} is overwhelmed by the noise that leads to an inferior binary result, whereas our method filters out these uncorrelated perturbations and delivers clean, recognizable letters with sharp boundaries. 

Fig.~\ref{fig:ex_quality}~(b) shows a case when sparse events from minimal motion are insufficient to complement structural cues, causing the degraded frame to dominate and yield suboptimal binarization (\eg, unclear ``X'' in ``EXIT''). For comparison, we also present an improved result at a later timestamp where stronger motion brings a denser event stream. To further support these visual findings, Fig.~\ref{fig:ex_quality}~(c) and~(d) examine our response to synthetic Poisson noise and random event dropout, evaluating six frames at different time within the same sequence under three lighting conditions. The variances indicate that while our method remains robust to moderate noise interference, it is not resilient to event scarcity in near-static scenes where neither modality conveys adequate discriminative features.

\input{src/deployment}
\subsection{Edge Deployment Verification}
Beyond the above evaluations on a desktop-class system, we further verify lightweight deployment on an edge platform, NVIDIA Jetson Orin Nano (\SI{8}{\giga\byte}, runs in CPU-only mode), to assess the practicability under real-world resource-constrained hardware. Table~\ref{tab:deployment} compares EDI (synchronous intensity reconstruction)~\cite{pan2019bringing}, CF (asynchronous intensity reconstruction)~\cite{scheerlinck2018continuous}, and ours (asynchronous binary-state propagation) at different frequencies. It shows the strong capability of our method to deliver both fast and accurate performance in edge deployment. More findings are that the asynchronous formulation, which benefits both runtime efficiency and reconstruction accuracy across varying frequencies, shows better scalability, while operating in the dense intensity domain often leads to a higher memory footprint redundant in binarization tasks. 

\section{Conclusion}\label{sec:end}
This work proposes a lightweight dual-modal method that bridges the features of frames and events to enable training-free, real-time, high-frame-rate binarization on CPU-only systems. Extensive experiments on real-world datasets demonstrate its competitive performance in mitigating motion blur and clear advantages under harsh illumination, with additional benefits for downstream tasks. 

Nevertheless, a known limitation remains. Our current formulation relies on global statistics and full-frame state updates for parameter estimation and binary state propagation. This design is a trade-off between robustness and efficiency, since global histogram and threshold statistics help preserve scene-level consistency and are less sensitive to local event noise or sparse observations. However, this also introduces avoidable computation compared with a fully local implementation. Particularly, some operations are applied over the entire spatial domain even though only few pixels are activated by events. A more fine-grained patch-wise or local update strategy could further reduce computational cost and improve scalability.

\appendices
\section{Quasi-Bimodality Justification}
\label{app:bimodal_statistics}
We give a simple statistical justification for the variance-matching assumption in Section~\ref{sec:c}. In quasi-bimodal scenes, photometric variation is dominated by foreground-background separation rather than by intra-class texture fluctuations. Consider an ideal bimodal log-radiance variable
\begin{equation}
    \ell(\mathbf{x},t) = \ln L(\mathbf{x},t) \in \{\ell_0,\ell_1\},
\end{equation}
where $\ell_0$ and $\ell_1$ denote the background and foreground levels, respectively. Let $\pi$ be the probability that a pixel belongs to the foreground class. Then, the mean log-radiance is
\begin{equation}
    \bar{\ell} = (1-\pi)\ell_0 + \pi\ell_1,
\end{equation}
and then the corresponding variance is
\begin{equation}
\begin{aligned}
    \sigma_\ell^2 &= (1-\pi)(\ell_0-\bar{\ell})^2 + \pi(\ell_1-\bar{\ell})^2  \\
&= \pi(1-\pi)(\ell_1-\ell_0)^2,
\end{aligned}
\end{equation}
which shows that the log-domain variance is mainly governed by the inter-class separation $\ell_1-\ell_0$ and the class balance $\pi(1-\pi)$ in an ideal bimodal scene. As such, the global dispersion is informative about foreground-background contrast. In real quasi-bimodal scenes, small intra-class fluctuations may exist due to textures, sensor noise, and illumination variations
\begin{equation}
    \ell(\mathbf{x},t) = z(\mathbf{x},t)\ell_1 + (1-z(\mathbf{x},t))\ell_0 + \eta(\mathbf{x},t),
\end{equation}
where $z(\mathbf{x},t)\in\{0,1\}$ denotes the class label and $\eta(\mathbf{x},t)$ is intra-class perturbation. When the perturbation is small compared with the foreground-background separation, the variance can be approximated as
\begin{equation}
    \sigma_\ell^2 \approx \pi(1-\pi)(\ell_1-\ell_0)^2 + \sigma_\eta^2.
\end{equation}
For quasi-bimodal objects, $\sigma_\eta^2$ is much smaller than the inter-class term, and $\sigma_\ell^2$ remains dominated by the binary structure.

While the frame and event accumulation are not identical physical quantities, their dispersions over $\mathbb{D}$ are both dominated by the same foreground-background transitions. This motivates the task-oriented moment matching in Eq.~\eqref{eq:contrast} to estimate an effective contrast scale for binarization, where reliable class separation matters over exact radiance reconstruction. 

\section{Exposure-Adaptive Fusion Solution}
We provide the derivation of exposure-adaptive fusion in Section~\ref{sec:fusion} 
The purpose of dynamic range reshaping is to construct a proxy frame that preserves reliable intensity information when the frame is trustworthy, while introducing event-driven structural cues when it is degraded by poor exposure or motion blur. Let $Q(\mathbf{x},t') = \mathbb{I}_{\mathbf{x} \in \mathbb{D}}\cdot \tilde{I}_\mathcal{E}(\mathbf{x},t') + \mathbb{I}_{\mathbf{x} \in \mathbb{S}} \cdot \mu$, then Eq.~\eqref{eq:fusion_min} is 
\begin{equation}
     I_\lambda = \arg\min_J \lambda \|J-I\|_2^2 +(1-\lambda)\|J-Q\|_2^2.
\end{equation}
Since the objective is separable over pixels, we can derive the solution pixel-wisely. For each pixel, the objective becomes
\begin{equation}
    \mathcal{L}(J) = \lambda\big(J(\mathbf{x})-I(\mathbf{x})\big)^2 + (1-\lambda)\big(J(\mathbf{x})-Q(\mathbf{x},t')\big)^2.
\end{equation}
Setting the derivative with respect to $J(\mathbf{x})$ to zero gives
\begin{equation}
    \frac{\partial \mathcal{L}}{\partial J(\mathbf{x})} = 2\lambda\big(J(\mathbf{x})-I(\mathbf{x})\big) + 2(1-\lambda)\big(J(\mathbf{x})-Q(\mathbf{x},t')\big) = 0,
\end{equation}
and rearranging the terms yields
\begin{equation}
    \lambda J(\mathbf{x}) + (1-\lambda)J(\mathbf{x}) = \lambda I(\mathbf{x}) + (1-\lambda)Q(\mathbf{x},t').
\end{equation}
Since $\lambda+(1-\lambda)=1$, the closed-form solution is
\begin{equation}
    J^*(\mathbf{x},t') = \lambda I(\mathbf{x}) + (1-\lambda)Q(\mathbf{x},t').
\end{equation}
Substituting $Q(\mathbf{x},t')$ gives
\begin{equation}
     I_\lambda(\mathbf{x},t') = \lambda I(\mathbf{x}) + (1-\lambda)\big(\mathbb{I}_{\mathbf{x} \in \mathbb{D}}\cdot \tilde{I}_\mathcal{E}(\mathbf{x},t') + \mathbb{I}_{\mathbf{x} \in \mathbb{S}} \cdot \mu\big).
\end{equation}
As such, the fusion improves histogram separability without iterative optimization or trainable parameters.

\section{Time and Space Complexity Analysis}
\input{src/complexity}
We offer a detailed breakdown of the computational complexity of each component of our method in Table~\ref{tab:complexity}. The analysis separates one-time initial binary frame estimation from high-frame-rate propagation, to highlight that the latter scales linearly with the number of events and remains independent of the requested output frame rate.

\section*{Acknowledgments}
We thank Rongzhou Chen (PhD student at HKU) for his valuable discussion and assistance, and the Imaging Systems Laboratory at HKU for supplying event cameras. This work was supported in part by the National Natural Science Foundation of China under Grant 62505167 and in part by the Science and Technology Commission of Shanghai Municipality under Grant 25ZR1402149.

\bibliographystyle{IEEEtran}
\bibliography{main_ref.bib}

\end{document}

%% file: src/intro.tex
\begin{figure}[t]
\centering

\subfloat[Edge detection by the traditional Sobel operator.]{
    \includegraphics[width=\linewidth]{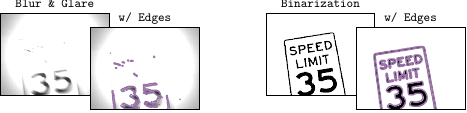}
}

\vspace{4pt}

\subfloat[Visual commonsense reasoning by modern MLLMs.]{
    \includegraphics[width=\linewidth]{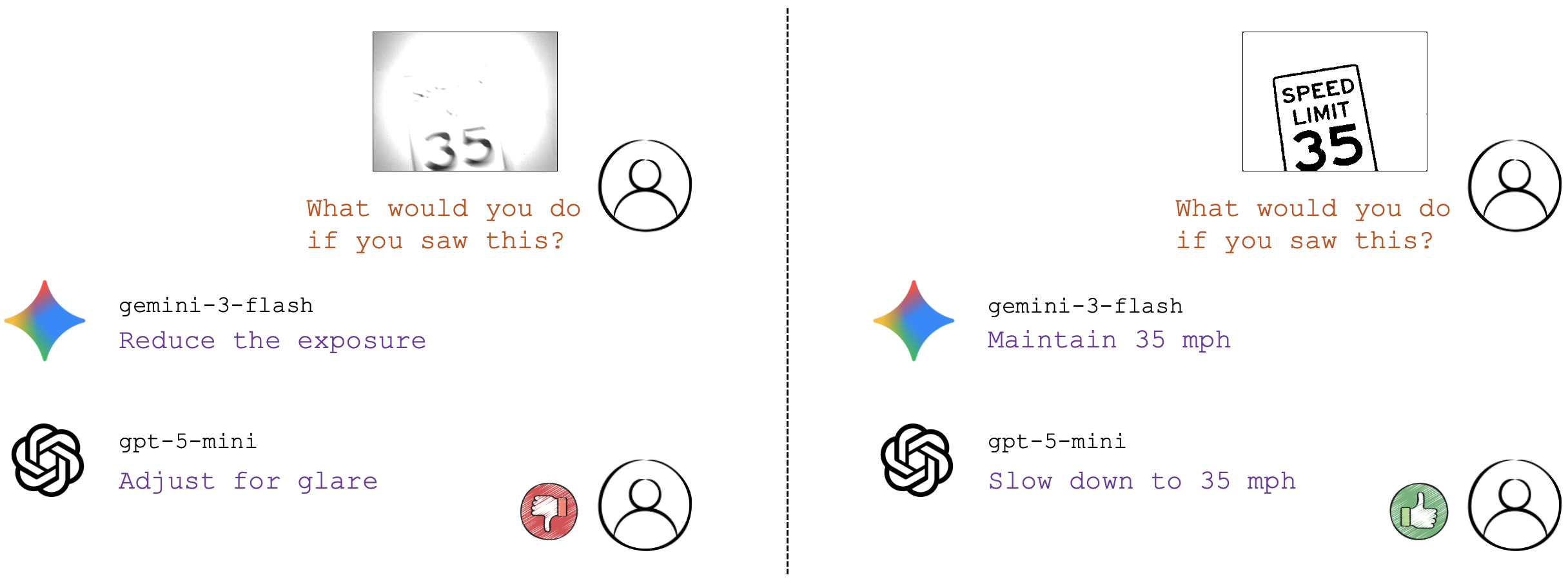}
}

\caption{Minimal input, maximum insight.}
\label{fig:intro}
\end{figure}

%% file: src/intro2.tex
\begin{figure}[t]
\centering
\includegraphics[width=\linewidth]{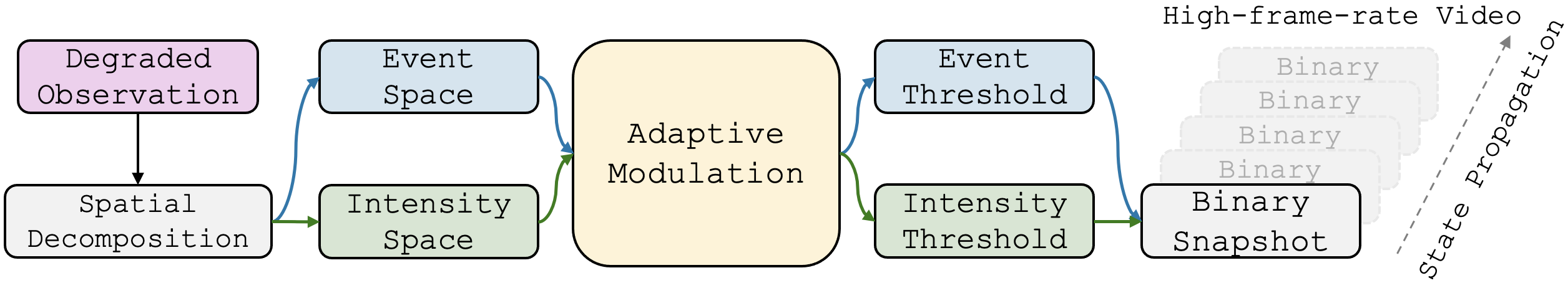}
\caption{Our unified training-free binarization pipeline receives degraded observations and outputs high-frame-rate binary videos.}
\label{fig:intro2}
\end{figure}

%% file: src/dual.tex
\begin{figure*}[!t]
\centering

\subfloat[Spatial domain $\mathrm{\Omega} = \mathbb{D} \cup \mathbb{S}$]{
    \begin{minipage}[b][2.5cm][t]{0.23\textwidth}
        \centering
        \includegraphics[width=\textwidth]{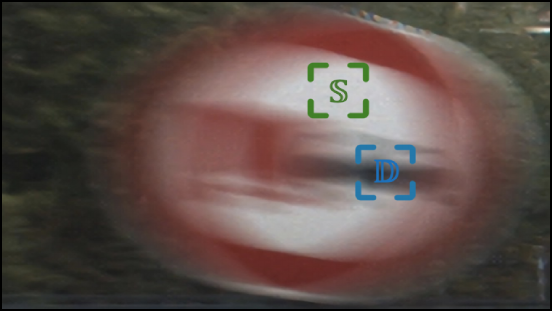}
    \end{minipage}
}\hfill
\subfloat[Inference in $\mathbb{S}$]{
    \begin{minipage}[b][2.5cm][t]{0.23\textwidth}
        \centering
        \includegraphics[width=\textwidth]{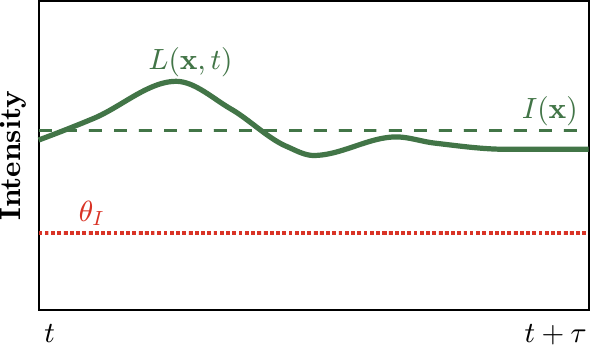}
    \end{minipage}
}\hfill
\subfloat[Degradation in $\mathbb{D}$]{
    \begin{minipage}[b][2.5cm][t]{0.23\textwidth}
        \centering
        \includegraphics[width=\textwidth]{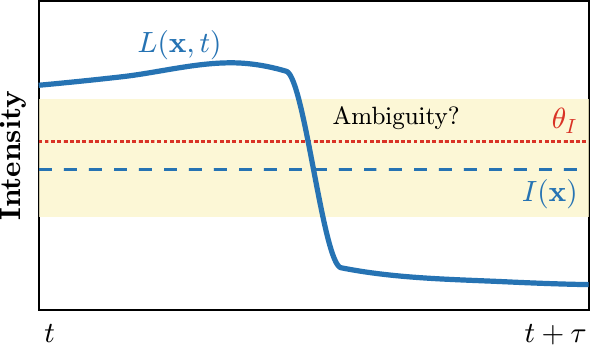}
    \end{minipage}
}\hfill
\subfloat[Inference in $\mathbb{D}$ by events]{
    \begin{minipage}[b][2.5cm][t]{0.23\textwidth}
        \centering
        \includegraphics[width=\textwidth]{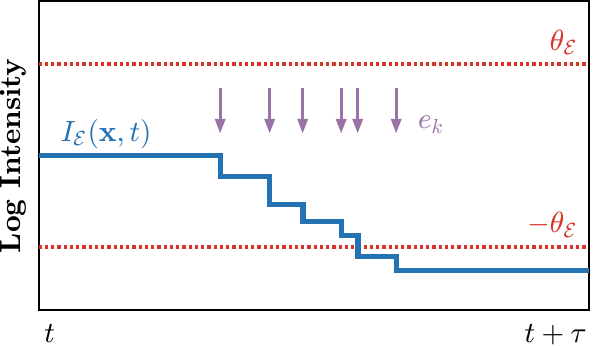}
    \end{minipage}
}

  \caption{The dual-modal binarization framework. (a) A blurry ``No overtaking'' sign observed during high-speed driving is spatially decomposed into $\mathbb{D}$ and $\mathbb{S}$. (b) Static pixels are faithfully binarized using $I(\mathbf{x})$ and $\theta_I$. (c) Motion blur in $\mathbb{D}$ pushes $I(\mathbf{x})$ into an ambiguity zone, causing intensity-based inference to fail. (d) Dynamic pixels are resolved by integrating events, where we decode a sharp binary state via $\pm\theta_\mathcal{E}$.}
  \label{fig:dual}
\end{figure*}

%% file: src/contrast.tex
\begin{figure}[!t]
  \centering
\subfloat[A quasi-bimodal OCR scene]{%
    \includegraphics[width=0.48\columnwidth]{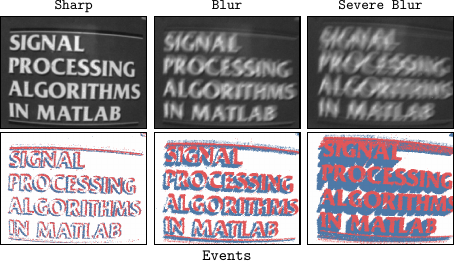}%
}
\hfill
\subfloat[Intensity distribution]{%
    \includegraphics[width=0.48\columnwidth]{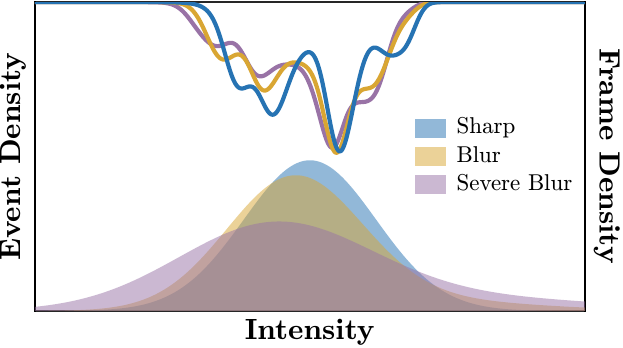}%
}
  \caption{Stronger motion brings heavier frame blur, yet the distribution in a quasi-bimodal scene remains preserved and informative (curves). Meanwhile, events increase and become more spatially dispersed along motion trajectories (shades). In both cases, structural responses are spread rather than concentrated, resulting in a reduced photometric variance.} 
  \label{fig:contrast}
\end{figure}

%% file: src/lambda.tex
\begin{figure}[!t]
  \centering
\subfloat[A quasi-bimodal OCR scene]{%
    \includegraphics[width=0.48\columnwidth]{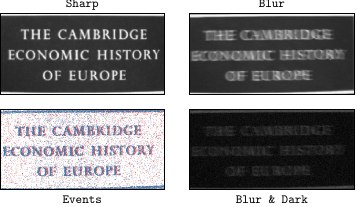}%
}
\hfill
\subfloat[Intensity distribution]{%
    \includegraphics[width=0.48\columnwidth]{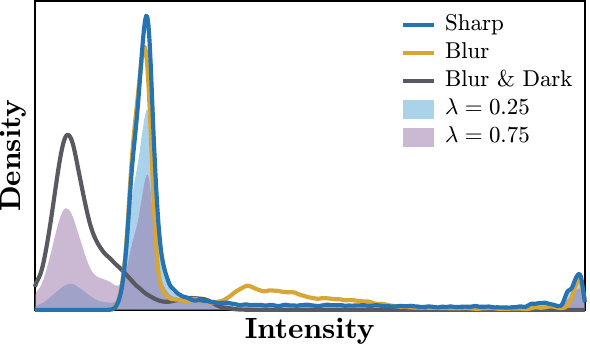}%
}
  \caption{A clear bimodal shape is observed in the sharp scene, while motion blur \& dark reduces class separability. We leverage the event prior ($\lambda = 0.25$) to recover missing structural details as well as a separable distribution.}
  \label{fig:lambda}
\end{figure}

%% file: src/limit.tex
\begin{figure}[!t]
  \centering
\subfloat[A quasi-bimodal OCR scene]{%
    \includegraphics[width=0.48\columnwidth]{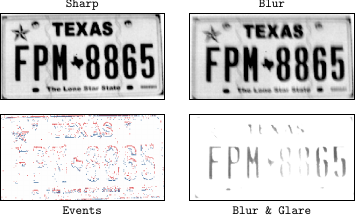}%
}
\hfill
\subfloat[Intensity distribution]{%
    \includegraphics[width=0.48\columnwidth]{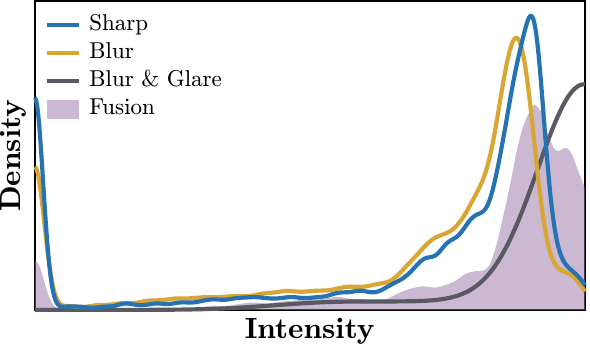}%
}
  \caption{A clear bimodal shape is observed in the sharp scene, while motion blur \& glare shows a unimodal distribution. Despite their HDR, event cameras respond to slight motion with limited events that fail to complement the saturated frame with meaningful motion cues, leading to ineffective fusion and a suboptimal distribution.}
  \label{fig:limit}
\end{figure}

%% file: src/algorithm.tex
\begin{algorithm}[t]
\caption{Asynchronous State Propagation}
\label{alg:asp}
\SetAlgoLined
\DontPrintSemicolon
\KwIn{$B(\mathbf{x}, t_0)$, $\mathcal{E}$, $c$, $\theta_{\mathcal{E}}$}
\KwOut{$B(\mathbf{x}, t)$}
Initialize $r(\mathbf{x}) \leftarrow 0$ for all $\mathbf{x}$\;
\ForEach{$e_k = (\mathbf{x}_k, t_k, p_k) \in \mathcal{E}$}{
    \If{$B(\mathbf{x}_k) \oplus \rho_k$}{
        $r(\mathbf{x}_k) \leftarrow r(\mathbf{x}_k) + p_k \cdot c$\;

        \If{$|r(\mathbf{x}_k)| \ge \theta_{\mathcal{E}}$}{
            $B(\mathbf{x}_k) \leftarrow 1 - B(\mathbf{x}_k)$\;
            $r(\mathbf{x}_k) \leftarrow 0$\;
$B(\mathbf{x}_k) \leftarrow \mathbb{I} \big[\sum_{\mathbf{x}' \in \mathcal{W}} B(\mathbf{x}') > \tfrac{1}{2} |\mathcal{W}| \big]$
        }
    }
}
\end{algorithm}

%% file: src/ex_binary_normal.tex
\begin{figure*}[!t]
    \centering
    \setlength{\fboxsep}{0pt}
    \setlength{\fboxrule}{0.5pt}
    
    \subfloat[Degradation]{%
        \begin{minipage}{0.120\linewidth}
            \begin{tikzpicture}
                \node[inner sep=0pt] (main1) {\fbox{\includegraphics[width=\dimexpr\linewidth-2\fboxrule\relax]{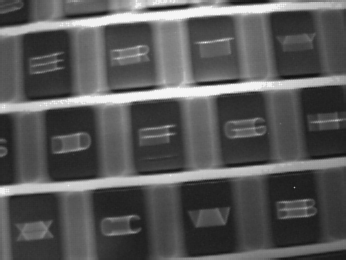}}};
                \node[inner sep=0pt, anchor=north east] at (main1.north east) {\fbox{\includegraphics[width=0.35\linewidth]{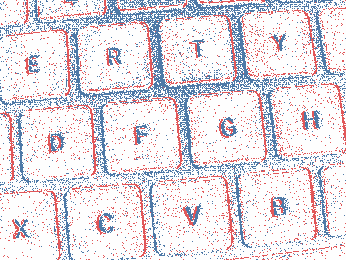}}};
            \end{tikzpicture}\\[3pt]
            \begin{tikzpicture}
                \node[inner sep=0pt] (main1) {\fbox{\includegraphics[width=\dimexpr\linewidth-2\fboxrule\relax]{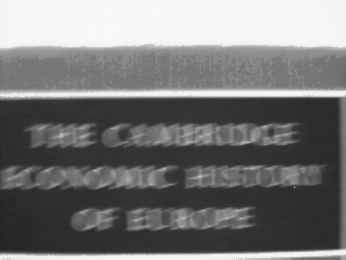}}};
                \node[inner sep=0pt, anchor=north east] at (main1.north east) {\fbox{\includegraphics[width=0.35\linewidth]{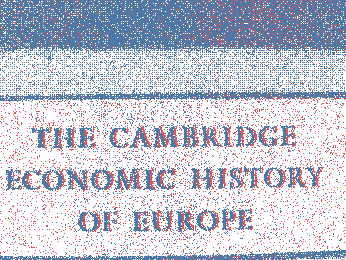}}};
            \end{tikzpicture}\\[3pt]
            \begin{tikzpicture}
                \node[inner sep=0pt] (main1) {\fbox{\includegraphics[width=\dimexpr\linewidth-2\fboxrule\relax]{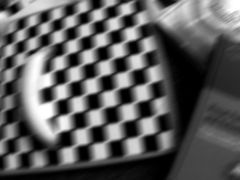}}};
                \node[inner sep=0pt, anchor=north east] at (main1.north east) {\fbox{\includegraphics[width=0.35\linewidth]{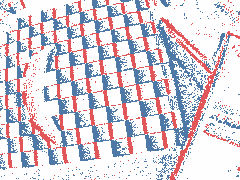}}};
            \end{tikzpicture}
        \end{minipage}%
    }\hfill
    \subfloat[Reference]{%
        \begin{minipage}{0.120\linewidth}
            \fbox{\includegraphics[width=\dimexpr\linewidth-2\fboxrule\relax]{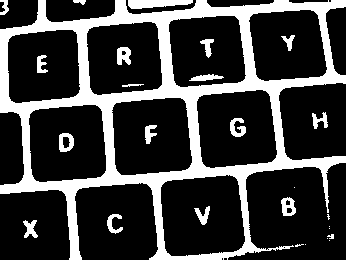}}\\[3pt]
            \fbox{\includegraphics[width=\dimexpr\linewidth-2\fboxrule\relax]{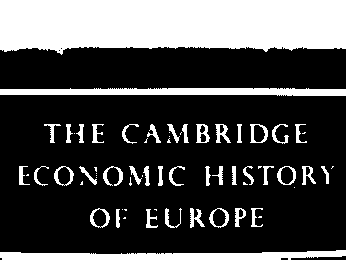}}\\[3pt]
            \fbox{\includegraphics[width=\dimexpr\linewidth-2\fboxrule\relax]{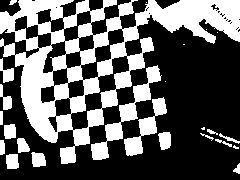}}
        \end{minipage}%
    }\hfill
    \subfloat[NICK~\cite{khurshid2009comparison}]{%
        \begin{minipage}{0.120\linewidth}
            \fbox{\includegraphics[width=\dimexpr\linewidth-2\fboxrule\relax]{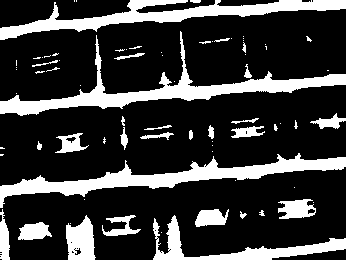}}\\[3pt]
            \fbox{\includegraphics[width=\dimexpr\linewidth-2\fboxrule\relax]{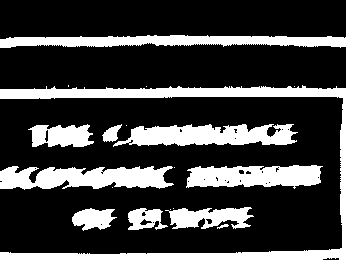}}\\[3pt]
            \fbox{\includegraphics[width=\dimexpr\linewidth-2\fboxrule\relax]{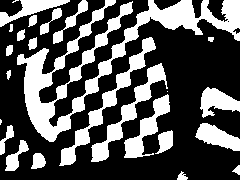}}
        \end{minipage}%
    }\hfill
    \subfloat[ISauvola~\cite{hadjadj2016isauvola}]{%
        \begin{minipage}{0.120\linewidth}
            \fbox{\includegraphics[width=\dimexpr\linewidth-2\fboxrule\relax]{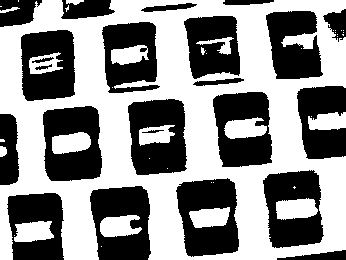}}\\[3pt]
            \fbox{\includegraphics[width=\dimexpr\linewidth-2\fboxrule\relax]{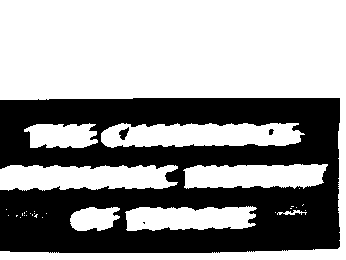}}\\[3pt]
            \fbox{\includegraphics[width=\dimexpr\linewidth-2\fboxrule\relax]{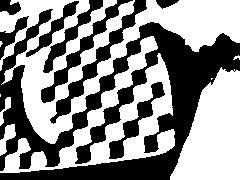}}
        \end{minipage}%
    }\hfill
    \subfloat[Wan~\cite{mustafa2018binarization}]{%
        \begin{minipage}{0.120\linewidth}
            \fbox{\includegraphics[width=\dimexpr\linewidth-2\fboxrule\relax]{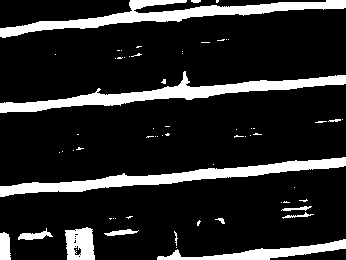}}\\[3pt]
            \fbox{\includegraphics[width=\dimexpr\linewidth-2\fboxrule\relax]{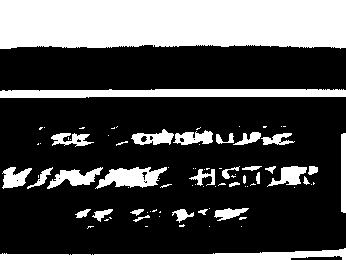}}\\[3pt]
            \fbox{\includegraphics[width=\dimexpr\linewidth-2\fboxrule\relax]{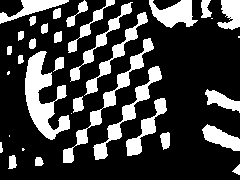}}
        \end{minipage}%
    }\hfill
    \subfloat[AE~\cite{calvo2019selectional}]{%
        \begin{minipage}{0.120\linewidth}
            \fbox{\includegraphics[width=\dimexpr\linewidth-2\fboxrule\relax]{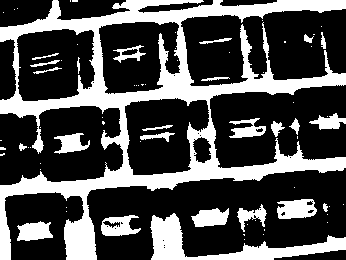}}\\[3pt]
            \fbox{\includegraphics[width=\dimexpr\linewidth-2\fboxrule\relax]{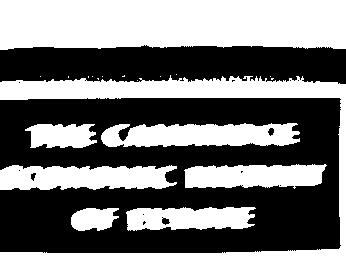}}\\[3pt]
            \fbox{\includegraphics[width=\dimexpr\linewidth-2\fboxrule\relax]{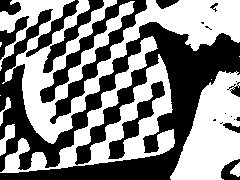}}
        \end{minipage}%
    }\hfill
    \subfloat[Lin~\cite{lin2024neuromorphic}]{%
        \begin{minipage}{0.120\linewidth}
            \fbox{\includegraphics[width=\dimexpr\linewidth-2\fboxrule\relax]{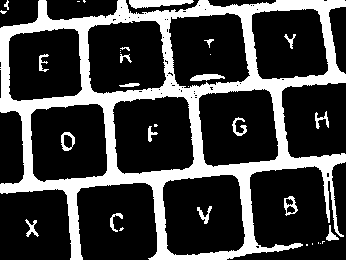}}\\[3pt]
            \fbox{\includegraphics[width=\dimexpr\linewidth-2\fboxrule\relax]{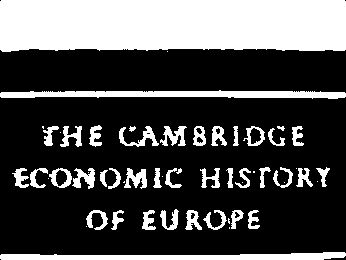}}\\[3pt]
            \fbox{\includegraphics[width=\dimexpr\linewidth-2\fboxrule\relax]{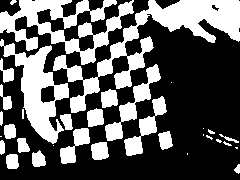}}
        \end{minipage}%
    }\hfill
    \subfloat[Ours]{%
        \begin{minipage}{0.120\linewidth}
            {\setlength{\fboxrule}{1pt}%
            \fcolorbox{boxred}{white}{\includegraphics[width=\dimexpr\linewidth-2\fboxrule\relax]{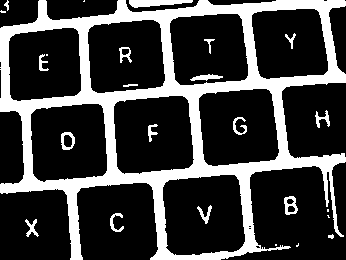}}}\\[3pt]
            {\setlength{\fboxrule}{1pt}%
            \fcolorbox{boxred}{white}{\includegraphics[width=\dimexpr\linewidth-2\fboxrule\relax]{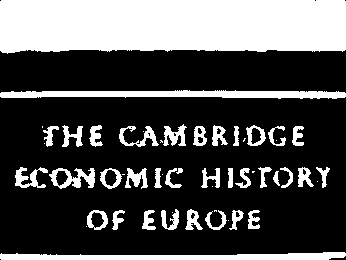}}}\\[3pt]
            {\setlength{\fboxrule}{1pt}%
            \fcolorbox{boxred}{white}{\includegraphics[width=\dimexpr\linewidth-2\fboxrule\relax]{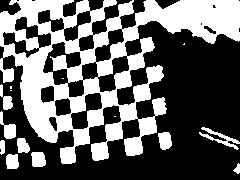}}}
        \end{minipage}%
    }

    \caption{Binarization results on the RND and REBlur datasets under motion blur. Each binary reference is taken from a sharp frame captured near the time of the degraded observation. The corresponding events are attached in (a) for reference.}
    \label{fig:ex_binary_normal}
\end{figure*}

%% file: src/ex_binary_low.tex
\begin{figure*}[!t]
    \centering
    \setlength{\fboxsep}{0pt}
    \setlength{\fboxrule}{0.5pt}
    
    \subfloat[Degradation]{%
        \begin{minipage}{0.120\linewidth}
            \begin{tikzpicture}
                \node[inner sep=0pt] (main2) {\fbox{\includegraphics[width=\dimexpr\linewidth-2\fboxrule\relax]{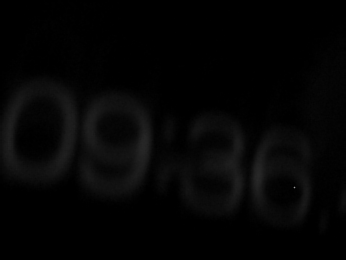}}};
                \node[inner sep=0pt, anchor=north east] at (main2.north east) {\fbox{\includegraphics[width=0.35\linewidth]{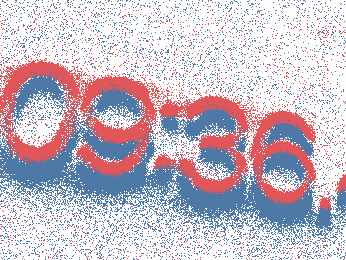}}};
            \end{tikzpicture}\\[3pt]
            \begin{tikzpicture}
                \node[inner sep=0pt] (main2) {\fbox{\includegraphics[width=\dimexpr\linewidth-2\fboxrule\relax]{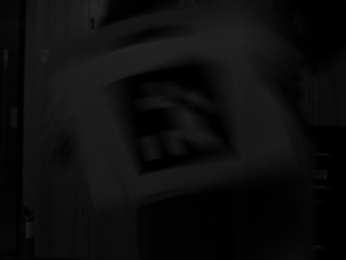}}};
                \node[inner sep=0pt, anchor=south west] at (main2.south west) {\fbox{\includegraphics[width=0.35\linewidth]{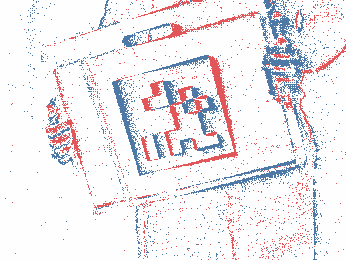}}};
            \end{tikzpicture}\\[3pt]
            \begin{tikzpicture}
                \node[inner sep=0pt] (main2) {\fbox{\includegraphics[width=\dimexpr\linewidth-2\fboxrule\relax]{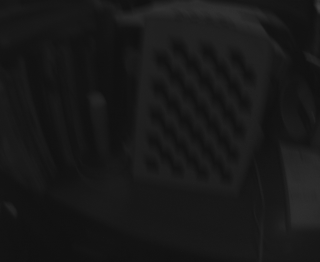}}};
                \node[inner sep=0pt, anchor=south west] at (main2.south west) {\fbox{\includegraphics[width=0.35\linewidth]{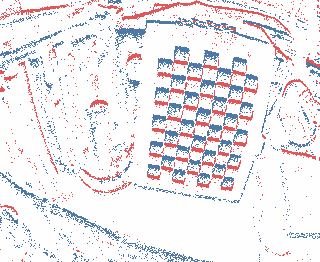}}};
            \end{tikzpicture}
        \end{minipage}%
    }\hfill
    \subfloat[Reference]{%
        \begin{minipage}{0.120\linewidth}
            \fbox{\includegraphics[width=\dimexpr\linewidth-2\fboxrule\relax]{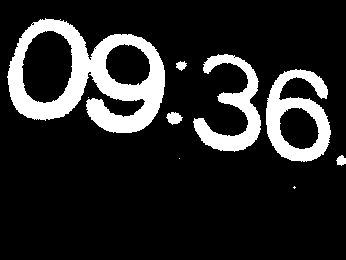}}\\[3pt]
            \fbox{\includegraphics[width=\dimexpr\linewidth-2\fboxrule\relax]{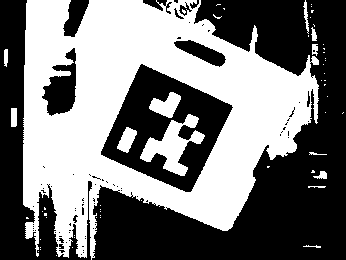}}\\[3pt]
            \fbox{\includegraphics[width=\dimexpr\linewidth-2\fboxrule\relax]{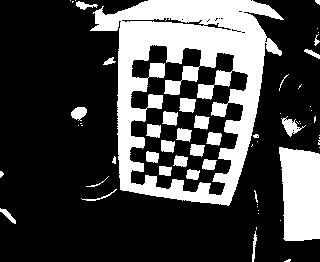}}
        \end{minipage}%
    }\hfill
    \subfloat[NICK~\cite{khurshid2009comparison}]{%
        \begin{minipage}{0.120\linewidth}
            \fbox{\includegraphics[width=\dimexpr\linewidth-2\fboxrule\relax]{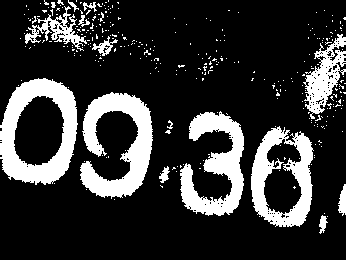}}\\[3pt]
            \fbox{\includegraphics[width=\dimexpr\linewidth-2\fboxrule\relax]{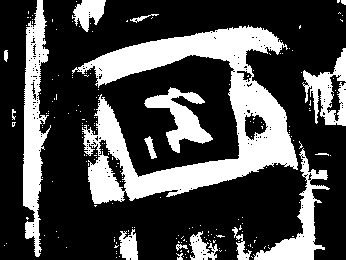}}\\[3pt]
            \fbox{\includegraphics[width=\dimexpr\linewidth-2\fboxrule\relax]{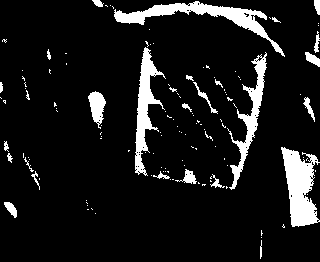}}
        \end{minipage}%
    }\hfill
    \subfloat[ISauvola~\cite{hadjadj2016isauvola}]{%
        \begin{minipage}{0.120\linewidth}
            \fbox{\includegraphics[width=\dimexpr\linewidth-2\fboxrule\relax]{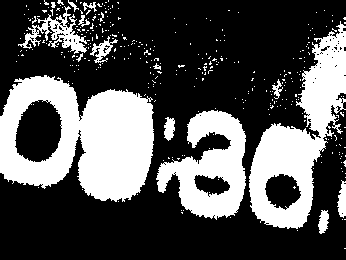}}\\[3pt]
            \fbox{\includegraphics[width=\dimexpr\linewidth-2\fboxrule\relax]{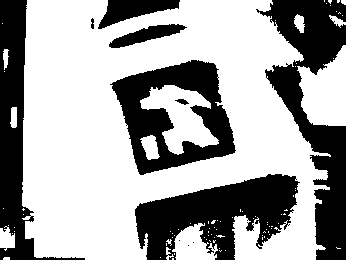}}\\[3pt]
            \fbox{\includegraphics[width=\dimexpr\linewidth-2\fboxrule\relax]{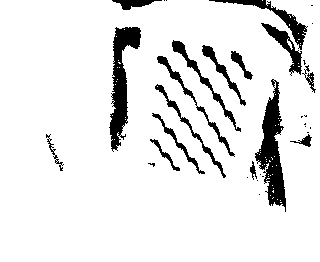}}
        \end{minipage}%
    }\hfill
    \subfloat[Wan~\cite{mustafa2018binarization}]{%
        \begin{minipage}{0.120\linewidth}
            \fbox{\includegraphics[width=\dimexpr\linewidth-2\fboxrule\relax]{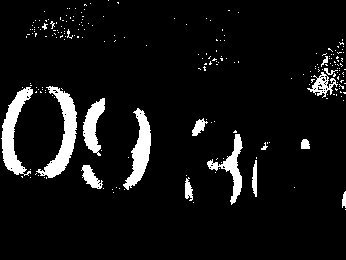}}\\[3pt]
            \fbox{\includegraphics[width=\dimexpr\linewidth-2\fboxrule\relax]{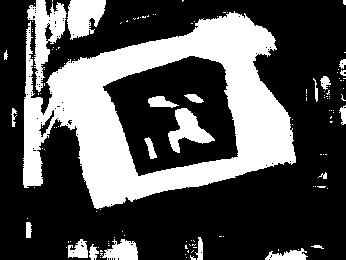}}\\[3pt]
            \fbox{\includegraphics[width=\dimexpr\linewidth-2\fboxrule\relax]{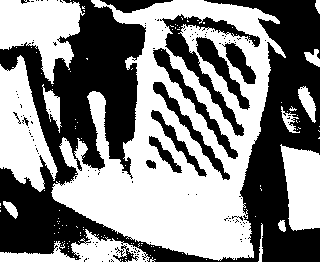}}
        \end{minipage}%
    }\hfill
    \subfloat[AE~\cite{calvo2019selectional}]{%
        \begin{minipage}{0.120\linewidth}
            \fbox{\includegraphics[width=\dimexpr\linewidth-2\fboxrule\relax]{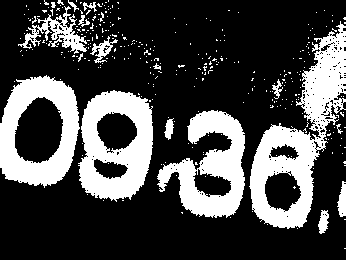}}\\[3pt]
            \fbox{\includegraphics[width=\dimexpr\linewidth-2\fboxrule\relax]{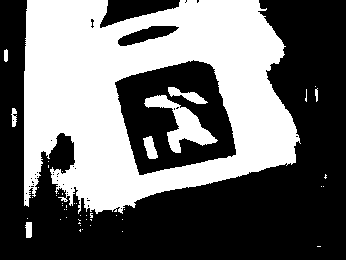}}\\[3pt]
            \fbox{\includegraphics[width=\dimexpr\linewidth-2\fboxrule\relax]{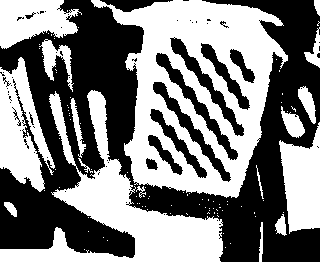}}
        \end{minipage}%
    }\hfill
    \subfloat[Lin~\cite{lin2024neuromorphic}]{%
        \begin{minipage}{0.120\linewidth}
            \fbox{\includegraphics[width=\dimexpr\linewidth-2\fboxrule\relax]{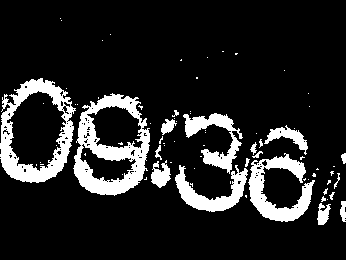}}\\[3pt]
            \fbox{\includegraphics[width=\dimexpr\linewidth-2\fboxrule\relax]{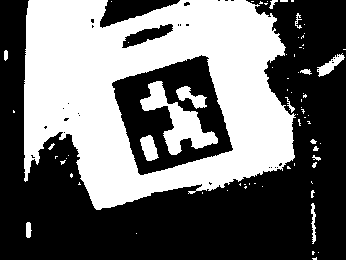}}\\[3pt]
            \fbox{\includegraphics[width=\dimexpr\linewidth-2\fboxrule\relax]{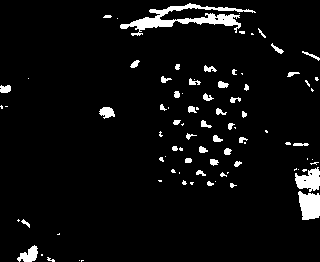}}
        \end{minipage}%
    }\hfill
    \subfloat[Ours]{%
        \begin{minipage}{0.120\linewidth}
            {\setlength{\fboxrule}{1pt}%
            \fcolorbox{boxred}{white}{\includegraphics[width=\dimexpr\linewidth-2\fboxrule\relax]{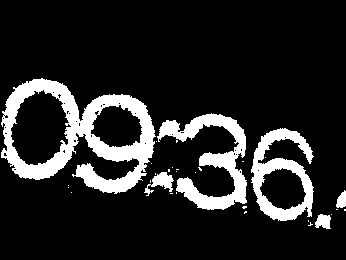}}}\\[3pt]
            {\setlength{\fboxrule}{1pt}%
            \fcolorbox{boxred}{white}{\includegraphics[width=\dimexpr\linewidth-2\fboxrule\relax]{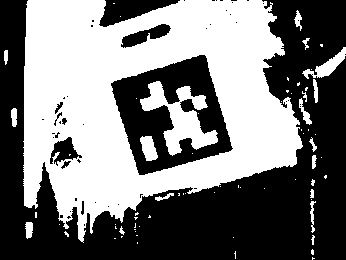}}}\\[3pt]
            {\setlength{\fboxrule}{1pt}%
            \fcolorbox{boxred}{white}{\includegraphics[width=\dimexpr\linewidth-2\fboxrule\relax]{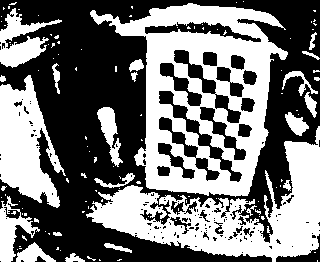}}}
        \end{minipage}%
    }
    
    \caption{Binarization results on the RND, REBlur, and EBT datasets under both low-light and motion-blur conditions. Each binary reference is taken from a sharp frame captured near the time of the degraded observation. The corresponding events are attached in (a) for reference.}
    \label{fig:ex_binary_low}
\end{figure*}

%% file: src/ex_binary_glare.tex
\begin{figure*}[!t]
    \centering
    \setlength{\fboxsep}{0pt}
    \setlength{\fboxrule}{0.5pt}
    
    \subfloat[Degradation]{%
        \begin{minipage}{0.120\linewidth}
            \begin{tikzpicture}
                \node[inner sep=0pt] (main3) {\fbox{\includegraphics[width=\dimexpr\linewidth-2\fboxrule\relax]{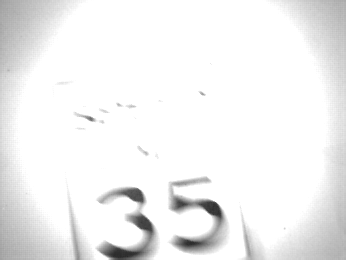}}};
                \node[inner sep=0pt, anchor=north east] at (main3.north east) {\fbox{\includegraphics[width=0.35\linewidth]{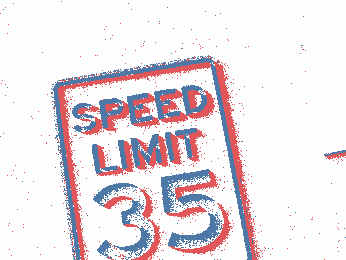}}};
            \end{tikzpicture}\\[3pt]
            \begin{tikzpicture}
                \node[inner sep=0pt] (main3) {\fbox{\includegraphics[width=\dimexpr\linewidth-2\fboxrule\relax]{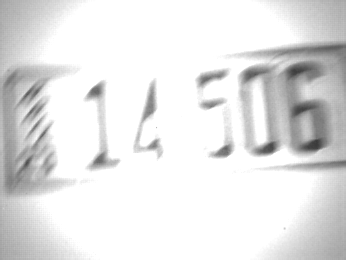}}};
                \node[inner sep=0pt, anchor=south east] at (main3.south east) {\fbox{\includegraphics[width=0.35\linewidth]{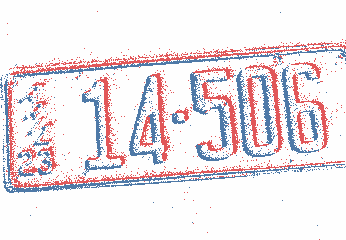}}};
            \end{tikzpicture}\\[3pt]
            \begin{tikzpicture}
                \node[inner sep=0pt] (main3) {\fbox{\includegraphics[width=\dimexpr\linewidth-2\fboxrule\relax]{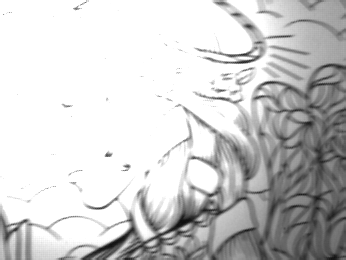}}};
                \node[inner sep=0pt, anchor=south east] at (main3.south east) {\fbox{\includegraphics[width=0.35\linewidth]{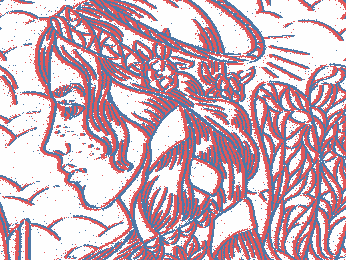}}};
            \end{tikzpicture}
        \end{minipage}%
    }\hfill
    \subfloat[Reference]{%
        \begin{minipage}{0.120\linewidth}
            \fbox{\includegraphics[width=\dimexpr\linewidth-2\fboxrule\relax]{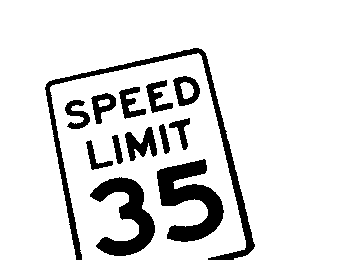}}\\[3pt]
            \fbox{\includegraphics[width=\dimexpr\linewidth-2\fboxrule\relax]{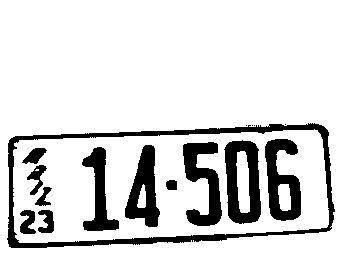}}\\[3pt]
            \fbox{\includegraphics[width=\dimexpr\linewidth-2\fboxrule\relax]{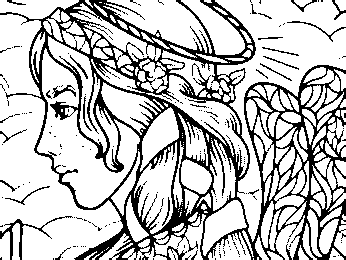}}
        \end{minipage}%
    }\hfill
    \subfloat[NICK~\cite{khurshid2009comparison}]{%
        \begin{minipage}{0.120\linewidth}
            \fbox{\includegraphics[width=\dimexpr\linewidth-2\fboxrule\relax]{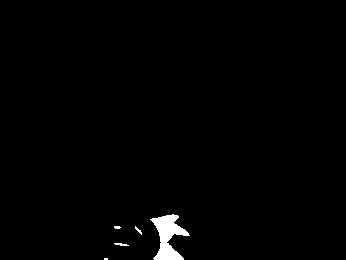}}\\[3pt]
            \fbox{\includegraphics[width=\dimexpr\linewidth-2\fboxrule\relax]{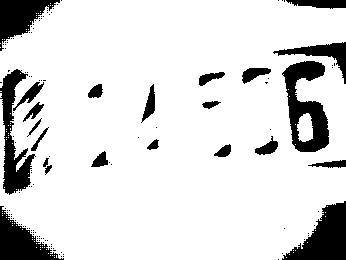}}\\[3pt]
            \fbox{\includegraphics[width=\dimexpr\linewidth-2\fboxrule\relax]{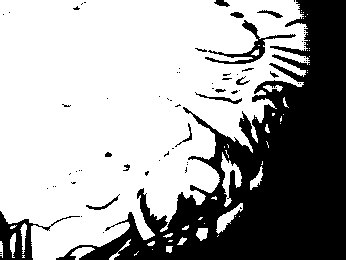}}
        \end{minipage}%
    }\hfill
    \subfloat[ISauvola~\cite{hadjadj2016isauvola}]{%
        \begin{minipage}{0.120\linewidth}
            \fbox{\includegraphics[width=\dimexpr\linewidth-2\fboxrule\relax]{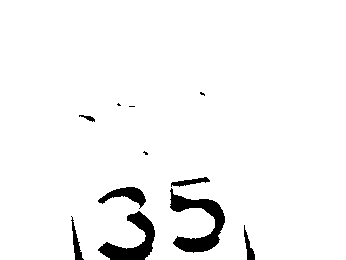}}\\[3pt]
            \fbox{\includegraphics[width=\dimexpr\linewidth-2\fboxrule\relax]{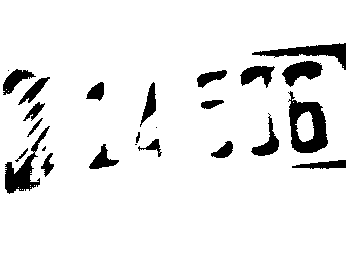}}\\[3pt]
            \fbox{\includegraphics[width=\dimexpr\linewidth-2\fboxrule\relax]{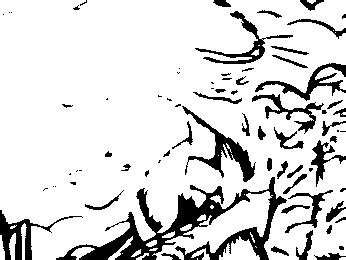}}
        \end{minipage}%
    }\hfill
    \subfloat[Wan~\cite{mustafa2018binarization}]{%
        \begin{minipage}{0.120\linewidth}
            \fbox{\includegraphics[width=\dimexpr\linewidth-2\fboxrule\relax]{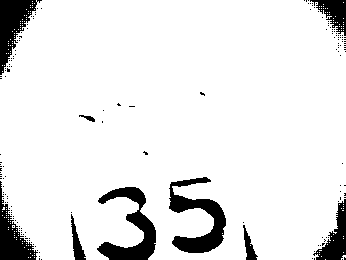}}\\[3pt]
            \fbox{\includegraphics[width=\dimexpr\linewidth-2\fboxrule\relax]{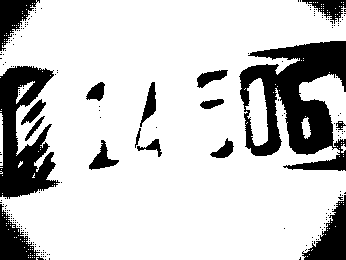}}\\[3pt]
            \fbox{\includegraphics[width=\dimexpr\linewidth-2\fboxrule\relax]{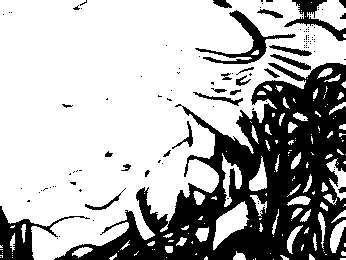}}
        \end{minipage}%
    }\hfill
    \subfloat[AE~\cite{calvo2019selectional}]{%
        \begin{minipage}{0.120\linewidth}
            \fbox{\includegraphics[width=\dimexpr\linewidth-2\fboxrule\relax]{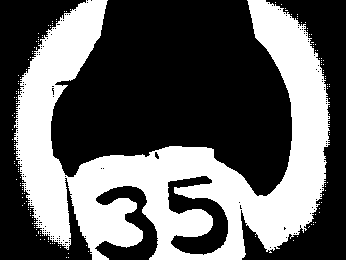}}\\[3pt]
            \fbox{\includegraphics[width=\dimexpr\linewidth-2\fboxrule\relax]{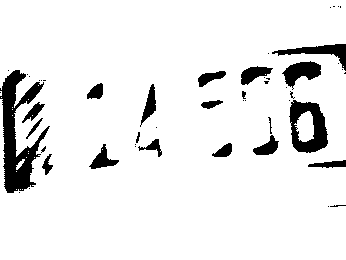}}\\[3pt]
            \fbox{\includegraphics[width=\dimexpr\linewidth-2\fboxrule\relax]{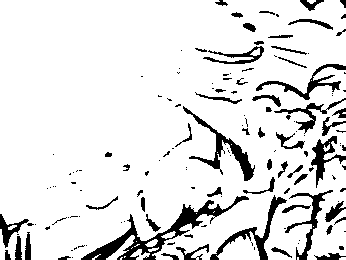}}
        \end{minipage}%
    }\hfill
    \subfloat[Lin~\cite{lin2024neuromorphic}]{%
        \begin{minipage}{0.120\linewidth}
            \fbox{\includegraphics[width=\dimexpr\linewidth-2\fboxrule\relax]{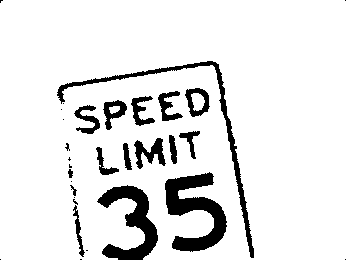}}\\[3pt]
            \fbox{\includegraphics[width=\dimexpr\linewidth-2\fboxrule\relax]{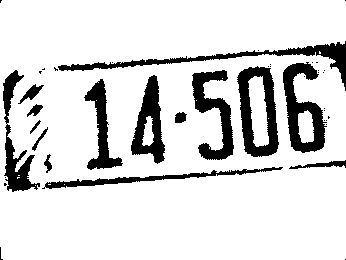}}\\[3pt]
            \fbox{\includegraphics[width=\dimexpr\linewidth-2\fboxrule\relax]{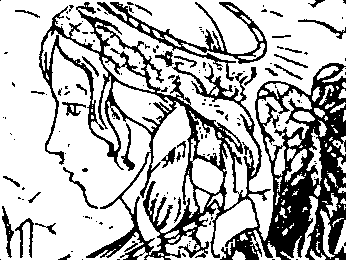}}
        \end{minipage}%
    }\hfill
    \subfloat[Ours]{%
        \begin{minipage}{0.120\linewidth}
            {\setlength{\fboxrule}{1pt}%
            \fcolorbox{boxred}{white}{\includegraphics[width=\dimexpr\linewidth-2\fboxrule\relax]{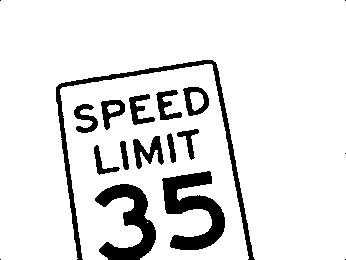}}}\\[3pt]
            {\setlength{\fboxrule}{1pt}%
            \fcolorbox{boxred}{white}{\includegraphics[width=\dimexpr\linewidth-2\fboxrule\relax]{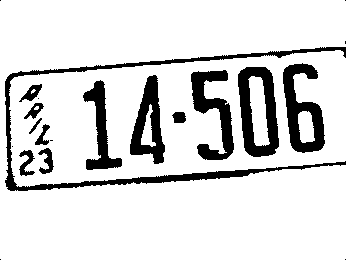}}}\\[3pt]
            {\setlength{\fboxrule}{1pt}%
            \fcolorbox{boxred}{white}{\includegraphics[width=\dimexpr\linewidth-2\fboxrule\relax]{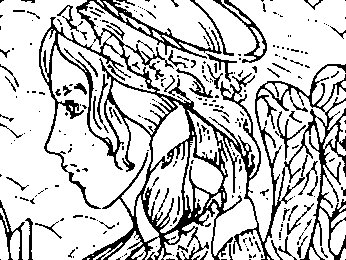}}}
        \end{minipage}%
    }
    
    \caption{Binarization results on the EBT dataset under both bright-glare and motion-blur conditions. Each binary reference is taken from a sharp frame captured near the time of the degraded observation. The corresponding events are attached in (a) for reference.}
    \label{fig:ex_binary_glare}
\end{figure*}

%% file: src/ex_reconstruction.tex
\begin{figure*}[!t]
    \centering
    \setlength{\fboxsep}{0pt}
    \setlength{\fboxrule}{0.5pt}
    
    \subfloat[Degradation]{%
        \begin{minipage}{0.120\linewidth}
            \begin{tikzpicture}
                \node[inner sep=0pt] (main1) {\fbox{\includegraphics[width=\dimexpr\linewidth-2\fboxrule\relax]{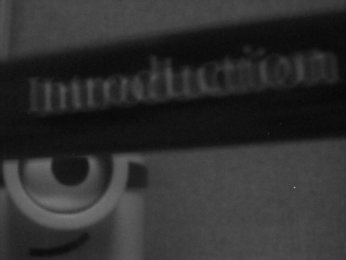}}};
                \node[inner sep=0pt, anchor=south east] at (main1.south east) {\fbox{\includegraphics[width=0.35\linewidth]{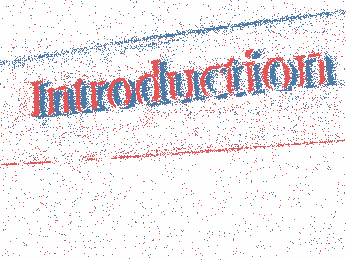}}};
            \end{tikzpicture}
            \\[3pt]
            \begin{tikzpicture}
                \node[inner sep=0pt] (main2) {\fbox{\includegraphics[width=\dimexpr\linewidth-2\fboxrule\relax]{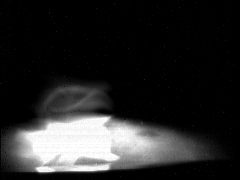}}};
                \node[inner sep=0pt, anchor=north east] at (main2.north east) {\fbox{\includegraphics[width=0.35\linewidth]{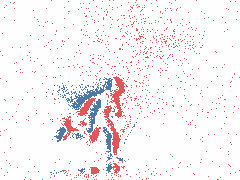}}};
            \end{tikzpicture}
        \end{minipage}%
    }\hfill
    \subfloat[Reference]{%
        \begin{minipage}{0.120\linewidth}
            \fbox{\includegraphics[width=\dimexpr\linewidth-2\fboxrule\relax]{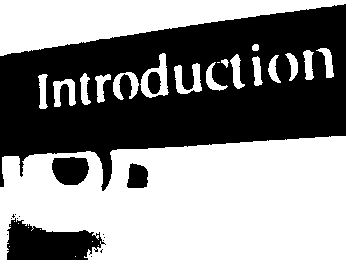}}
            \\[3pt]
            \fbox{\includegraphics[width=\dimexpr\linewidth-2\fboxrule\relax]{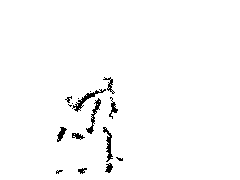}}
        \end{minipage}%
    }\hfill
    \subfloat[Bai~\cite{bai2018graph}]{%
        \begin{minipage}{0.120\linewidth}
            \begin{tikzpicture}
                \node[inner sep=0pt] (main2) {\fbox{\includegraphics[width=\dimexpr\linewidth-2\fboxrule\relax]{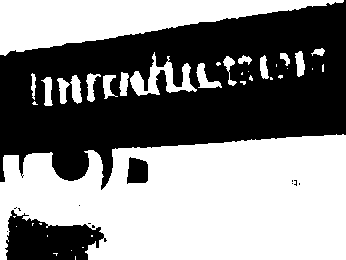}}};
                \node[inner sep=0pt, anchor=south east] at (main2.south east) {\fcolorbox{boxblue}{white}{\includegraphics[width=0.35\linewidth]{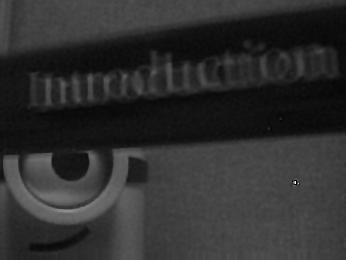}}};
            \end{tikzpicture}
            \\[3pt]
            \begin{tikzpicture}
                \node[inner sep=0pt] (main2) {\fbox{\includegraphics[width=\dimexpr\linewidth-2\fboxrule\relax]{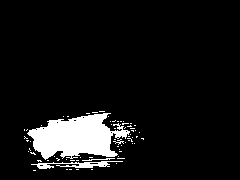}}};
                \node[inner sep=0pt, anchor=north east] at (main2.north east) {\fcolorbox{boxblue}{white}{\includegraphics[width=0.35\linewidth]{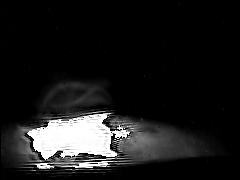}}};
            \end{tikzpicture}
        \end{minipage}%
    }\hfill
    \subfloat[Chen~\cite{chen2019blind}]{%
        \begin{minipage}{0.120\linewidth}
            \begin{tikzpicture}
                \node[inner sep=0pt] (main2) {\fbox{\includegraphics[width=\dimexpr\linewidth-2\fboxrule\relax]{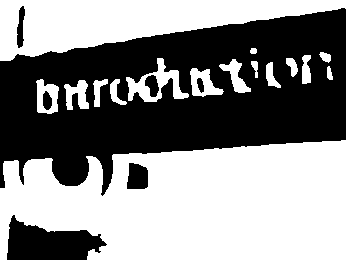}}};
                \node[inner sep=0pt, anchor=south east] at (main2.south east) {\fcolorbox{boxblue}{white}{\includegraphics[width=0.35\linewidth]{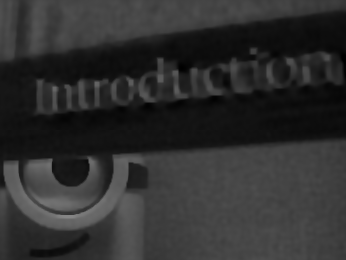}}};
            \end{tikzpicture}
            \\[3pt]
            \begin{tikzpicture}
                \node[inner sep=0pt] (main2) {\fbox{\includegraphics[width=\dimexpr\linewidth-2\fboxrule\relax]{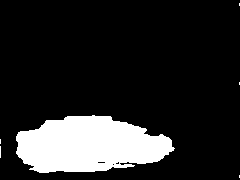}}};
                \node[inner sep=0pt, anchor=north east] at (main2.north east) {\fcolorbox{boxblue}{white}{\includegraphics[width=0.35\linewidth]{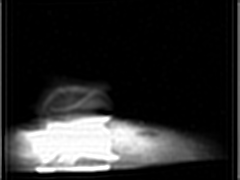}}};
            \end{tikzpicture}
        \end{minipage}%
    }\hfill
    \subfloat[CF~\cite{scheerlinck2018continuous}]{%
        \begin{minipage}{0.120\linewidth}
            \begin{tikzpicture}
                \node[inner sep=0pt] (main2) {\fbox{\includegraphics[width=\dimexpr\linewidth-2\fboxrule\relax]{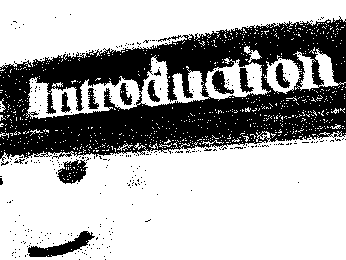}}};
                \node[inner sep=0pt, anchor=south east] at (main2.south east) {\fcolorbox{boxblue}{white}{\includegraphics[width=0.35\linewidth]{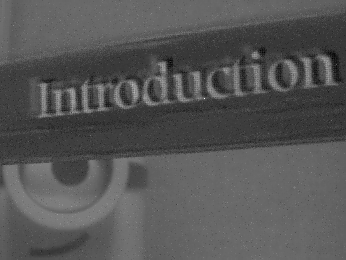}}};
            \end{tikzpicture}
            \\[3pt]
            \begin{tikzpicture}
                \node[inner sep=0pt] (main2) {\fbox{\includegraphics[width=\dimexpr\linewidth-2\fboxrule\relax]{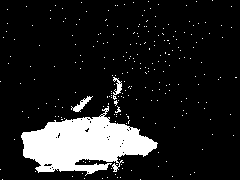}}};
                \node[inner sep=0pt, anchor=north east] at (main2.north east) {\fcolorbox{boxblue}{white}{\includegraphics[width=0.35\linewidth]{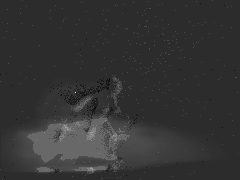}}};
            \end{tikzpicture}
        \end{minipage}%
    }\hfill
    \subfloat[EDI~\cite{pan2019bringing}]{%
        \begin{minipage}{0.120\linewidth}
            \begin{tikzpicture}
                \node[inner sep=0pt] (main2) {\fbox{\includegraphics[width=\dimexpr\linewidth-2\fboxrule\relax]{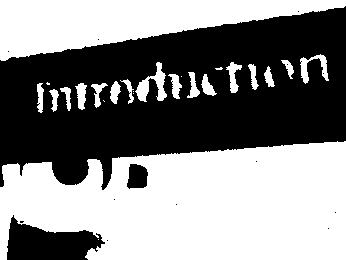}}};
                \node[inner sep=0pt, anchor=south east] at (main2.south east) {\fcolorbox{boxblue}{white}{\includegraphics[width=0.35\linewidth]{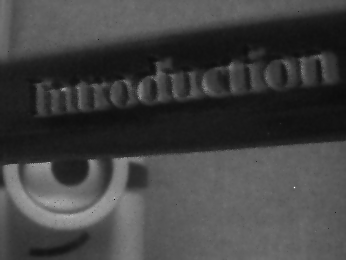}}};
            \end{tikzpicture}
            \\[3pt]
            \begin{tikzpicture}
                \node[inner sep=0pt] (main2) {\fbox{\includegraphics[width=\dimexpr\linewidth-2\fboxrule\relax]{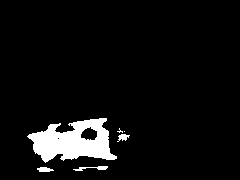}}};
                \node[inner sep=0pt, anchor=north east] at (main2.north east) {\fcolorbox{boxblue}{white}{\includegraphics[width=0.35\linewidth]{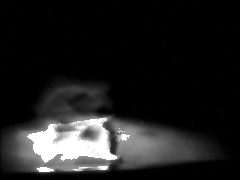}}};
            \end{tikzpicture}
        \end{minipage}%
    }\hfill
    \subfloat[Zhang~\cite{zhang2024tip}]{%
        \begin{minipage}{0.120\linewidth}
            \begin{tikzpicture}
                \node[inner sep=0pt] (main2) {\fbox{\includegraphics[width=\dimexpr\linewidth-2\fboxrule\relax]{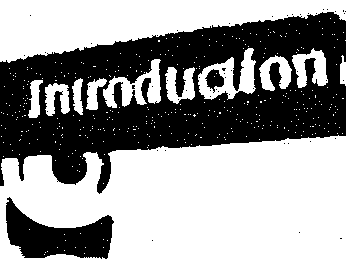}}};
                \node[inner sep=0pt, anchor=south east] at (main2.south east) {\fcolorbox{boxblue}{white}{\includegraphics[width=0.35\linewidth]{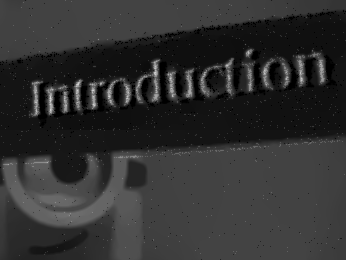}}};
            \end{tikzpicture}\\[3pt]
            \begin{tikzpicture}
                \node[inner sep=0pt] (main2) {\fbox{\includegraphics[width=\dimexpr\linewidth-2\fboxrule\relax]{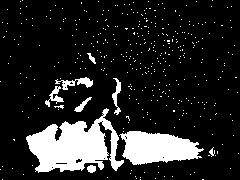}}};
                \node[inner sep=0pt, anchor=north east] at (main2.north east) {\fcolorbox{boxblue}{white}{\includegraphics[width=0.35\linewidth]{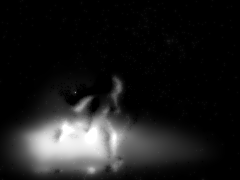}}};
            \end{tikzpicture}
        \end{minipage}%
    }\hfill
    \subfloat[Ours]{%
        \begin{minipage}{0.120\linewidth}
            {\setlength{\fboxrule}{1pt}%
            \fcolorbox{boxred}{white}{\includegraphics[width=\dimexpr\linewidth-2\fboxrule\relax]{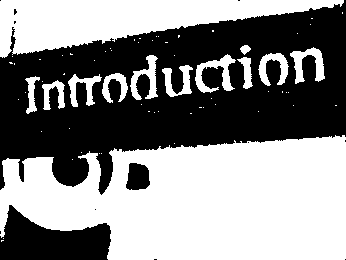}}}\\[3pt]
            {\setlength{\fboxrule}{1pt}%
            \fcolorbox{boxred}{white}{\includegraphics[width=\dimexpr\linewidth-2\fboxrule\relax]{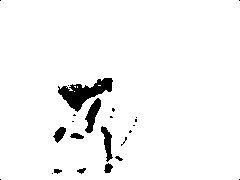}}}
        \end{minipage}%
    }
    
    \caption{Binarization results on the RND and CF datasets. Each binary reference is taken from a sharp frame captured near the time of the degraded observation. The corresponding events as well as the motion-deblurred intensity results are attached for reference.}
    \label{fig:ex_reconstruction}
\end{figure*}

%% file: src/quan_comparison.tex
\begin{table}[t]
\centering
\caption{Quantitative binarization results (with the \best{best} and \second{runner-up}), reported as \ding{172}~MCC~$\uparrow$~\ding{173}~PSNR~$\uparrow$~\ding{174}~NRM~$\downarrow$.}
\label{tab:quan_comparison}
\resizebox{\columnwidth}{!}{
\begin{tabular}{l c ccc}
\toprule
\multirow{2}{*}{Method} & Dual & Gentle Light & Low Light & Bright Glare \\
\cmidrule(lr){3-3} \cmidrule(lr){4-4} \cmidrule(lr){5-5}
 & Modal & \ding{172} / \ding{173} / \ding{174} & \ding{172} / \ding{173} / \ding{174} & \ding{172} / \ding{173} / \ding{174} \\
\midrule
\multicolumn{5}{c}{\textbf{HQF}} \\
\midrule
NICK~\cite{khurshid2009comparison}      &   & 0.31 / 4.15 / 0.40 & 0.09 / 1.73 / 0.45 & 0.13 / 2.43 / 0.40 \\
ISauvola~\cite{hadjadj2016isauvola}  &   & 0.34 / 5.03 / 0.36 & 0.12 / 1.75 / 0.43 & 0.17 / 2.89 / 0.41 \\
Wan~\cite{mustafa2018binarization}       &   & 0.35 / 5.37 / 0.33 & 0.14 / 2.10 / 0.39 & 0.14 / 2.67 / 0.38 \\
AE~\cite{calvo2019selectional}        &   & 0.35 / 5.35 / 0.34 & 0.15 / 2.26 / 0.38 & 0.13 / 2.54 / 0.40 \\
Lin~\cite{lin2024neuromorphic}       & $\checkmark$ & \best{0.62} / \second{6.91} / \best{0.16} & \second{0.31} / \second{4.36} / 0.36 & \second{0.49} / \second{6.08} / \second{0.29} \\
\cmidrule{1-5}
Bai~\cite{bai2018graph}        &   & 0.23 / 3.88 / 0.43 & 0.08 / 1.61 / 0.47 & 0.10 / 1.94 / 0.40 \\
Chen~\cite{chen2019blind}       &   & 0.20 / 3.85 / 0.44 & 0.10 / 1.69 / 0.47 & 0.11 / 2.01 / 0.41 \\
CF~\cite{scheerlinck2018continuous}        & $\checkmark$ & 0.48 / 5.54 / 0.30 & 0.23 / 3.15 / \second{0.33} & 0.32 / 5.04 / 0.31 \\
EDI~\cite{pan2019bringing}         & $\checkmark$ & 0.37 / 5.13 / 0.33 & 0.20 / 2.34 / 0.39 & 0.33 / 4.96 / 0.31 \\
Zhang~\cite{zhang2024tip}     & $\checkmark$ & \second{0.52} / 6.02 / 0.24 & 0.23 / 3.66 / 0.36 & 0.35 / 5.10 / 0.33 \\
\cmidrule{1-5}
\textbf{Ours} & $\checkmark$ & \best{0.62} / \best{6.93} / \second{0.21} & \best{0.48} / \best{6.21} / \best{0.26} & \best{0.54} / \best{6.60} / \best{0.25} \\
\midrule
\multicolumn{5}{c}{\textbf{REBlur}} \\
\midrule
NICK~\cite{khurshid2009comparison}      &   & 0.34 / 6.13 / 0.35 & 0.10 / 3.31 / 0.40 & 0.16 / 4.55 / 0.35 \\
ISauvola~\cite{hadjadj2016isauvola}  &   & 0.42 / 6.25 / 0.31 & 0.08 / 2.94 / 0.42 & 0.20 / 4.97 / 0.34 \\
Wan~\cite{mustafa2018binarization}       &   & 0.49 / 7.56 / 0.27 & 0.17 / 3.25 / 0.39 & 0.19 / 4.56 / 0.34 \\
AE~\cite{calvo2019selectional}        &   & 0.37 / 6.30 / 0.32 & 0.22 / 3.46 / 0.40 & 0.12 / 4.01 / 0.41 \\
Lin~\cite{lin2024neuromorphic}       & $\checkmark$ & \second{0.83} / \second{13.01} / \second{0.10} & \second{0.42} / \second{6.55} / \second{0.30} & \second{0.50} / \second{7.59} / \second{0.28} \\
\cmidrule{1-5}
Bai~\cite{bai2018graph}       &   & 0.40 / 6.21 / 0.30 & 0.22 / 3.59 / 0.44 & 0.14 / 4.22 / 0.41 \\
Chen~\cite{chen2019blind}       &   & 0.43 / 6.32 / 0.28 & 0.23 / 3.73 / 0.38 & 0.18 / 4.90 / 0.36 \\
CF~\cite{scheerlinck2018continuous}         & $\checkmark$ & 0.59 / 8.08 / 0.19 & 0.34 / 5.94 / 0.33 & 0.40 / 6.86 / 0.31 \\
EDI~\cite{pan2019bringing}         & $\checkmark$ & 0.67 / 8.77 / 0.14 & 0.35 / 6.12 / 0.32 & 0.43 / 7.29 / \second{0.28} \\
Zhang~\cite{zhang2024tip}      & $\checkmark$ & 0.71 / 10.96 / 0.12 & 0.30 / 5.79 / 0.36 & 0.39 / 6.22 / 0.35 \\
\cmidrule{1-5}
\textbf{Ours} & $\checkmark$ & \best{0.86} / \best{13.82} / \best{0.08} & \best{0.65} / \best{8.91} / \best{0.20} & \best{0.59} / \best{8.13} / \best{0.24} \\
\midrule
\multicolumn{5}{c}{\textbf{EBT}} \\
\midrule
NICK~\cite{khurshid2009comparison}      &   & 0.28 / 6.24 / 0.40 & 0.07 / 2.88 / 0.47 & 0.10 / 2.96 / 0.45 \\
ISauvola~\cite{hadjadj2016isauvola}  &   & 0.25 / 6.18 / 0.39 & 0.12 / 3.12 / 0.45 & 0.11 / 3.11 / 0.42 \\
Wan~\cite{mustafa2018binarization}       &   & 0.31 / 7.12 / 0.37 & 0.15 / 3.79 / 0.44 & 0.17 / 4.12 / 0.37 \\
AE~\cite{calvo2019selectional}        &   & 0.22 / 6.04 / 0.42 & 0.19 / 4.01 / 0.39 & 0.15 / 3.97 / 0.37 \\
Lin~\cite{lin2024neuromorphic}       & $\checkmark$ & \best{0.79} / \best{14.01} / \second{0.20} & \second{0.43} / \second{7.01} / \second{0.32} & \second{0.59} / \second{10.11} / \second{0.27} \\
\cmidrule{1-5}
Bai~\cite{bai2018graph}        &   & 0.30 / 7.19 / 0.37 & 0.18 / 3.95 / 0.39 & 0.20 / 4.15 / 0.39 \\
Chen~\cite{chen2019blind}      &   & 0.28 / 6.13 / 0.39 & 0.14 / 3.70 / 0.44 & 0.19 / 4.23 / 0.35 \\
CF~\cite{scheerlinck2018continuous}         & $\checkmark$ & 0.54 / 9.53 / 0.33 & 0.27 / 5.96 / 0.35 & 0.30 / 6.32 / 0.38 \\
EDI~\cite{pan2019bringing}       & $\checkmark$ & 0.59 / 10.01 / 0.28 & 0.29 / 6.15 / 0.36 & 0.35 / 7.10 / 0.35 \\
Zhang~\cite{zhang2024tip}      & $\checkmark$ & 0.65 / 11.79 / 0.24 & 0.32 / 6.93 / 0.33 & 0.41 / 7.25 / 0.34 \\
\cmidrule{1-5}
\textbf{Ours} & $\checkmark$ & \second{0.73} / \second{13.67} / \best{0.18} & \best{0.54} / \best{7.32} / \best{0.29} & \best{0.66} / \best{10.53} / \best{0.22} \\
\bottomrule
\end{tabular}
}
\end{table}

%% file: src/ex_downstream.tex
\begin{figure}[!t]
  \centering
  \setlength{\fboxsep}{0pt}
  \setlength{\fboxrule}{0.5pt}
  
  \subfloat[OCR for the text ``MSA Camps Events'']{%
    \begin{minipage}{\columnwidth}
      \centering
      \fbox{\includegraphics[width=0.23\linewidth]{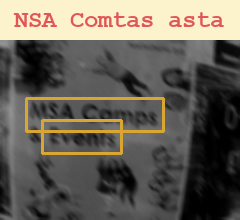}}
      \fbox{\includegraphics[width=0.23\linewidth]{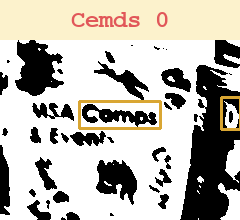}}
      \fbox{\includegraphics[width=0.23\linewidth]{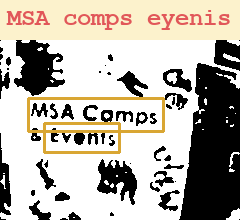}}
      \fbox{\includegraphics[width=0.23\linewidth]{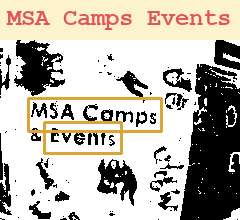}}
    \end{minipage}%
  } \\[6pt]  

  \subfloat[Fiducial Marker Tracking]{%
    \begin{minipage}{\columnwidth}
      \centering
      \fbox{\includegraphics[width=0.23\linewidth]{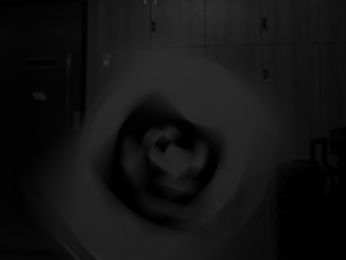}}
      \fbox{\includegraphics[width=0.23\linewidth]{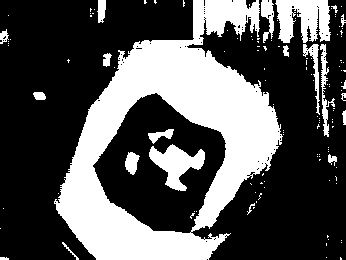}}
      \fbox{\includegraphics[width=0.23\linewidth]{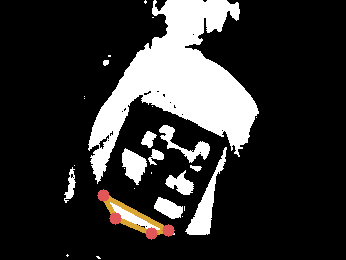}}
      \fbox{\includegraphics[width=0.23\linewidth]{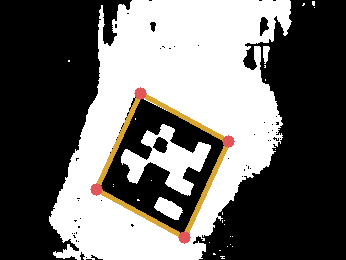}}
    \end{minipage}%
  } \\[6pt] 

  \subfloat[Optical Flow Estimation]{%
    \begin{minipage}{\columnwidth}
      \centering
      \begin{tikzpicture}[baseline=(main1.base)] 
        \node[inner sep=0pt] (main1) {
        \fbox{\includegraphics[width=0.23\linewidth]{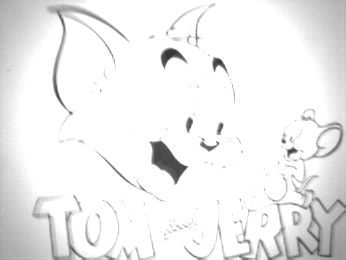}}
        };
        \node[inner sep=0pt, anchor=north east] at (main1.north east) {
            \fbox{\includegraphics[width=0.05\linewidth]{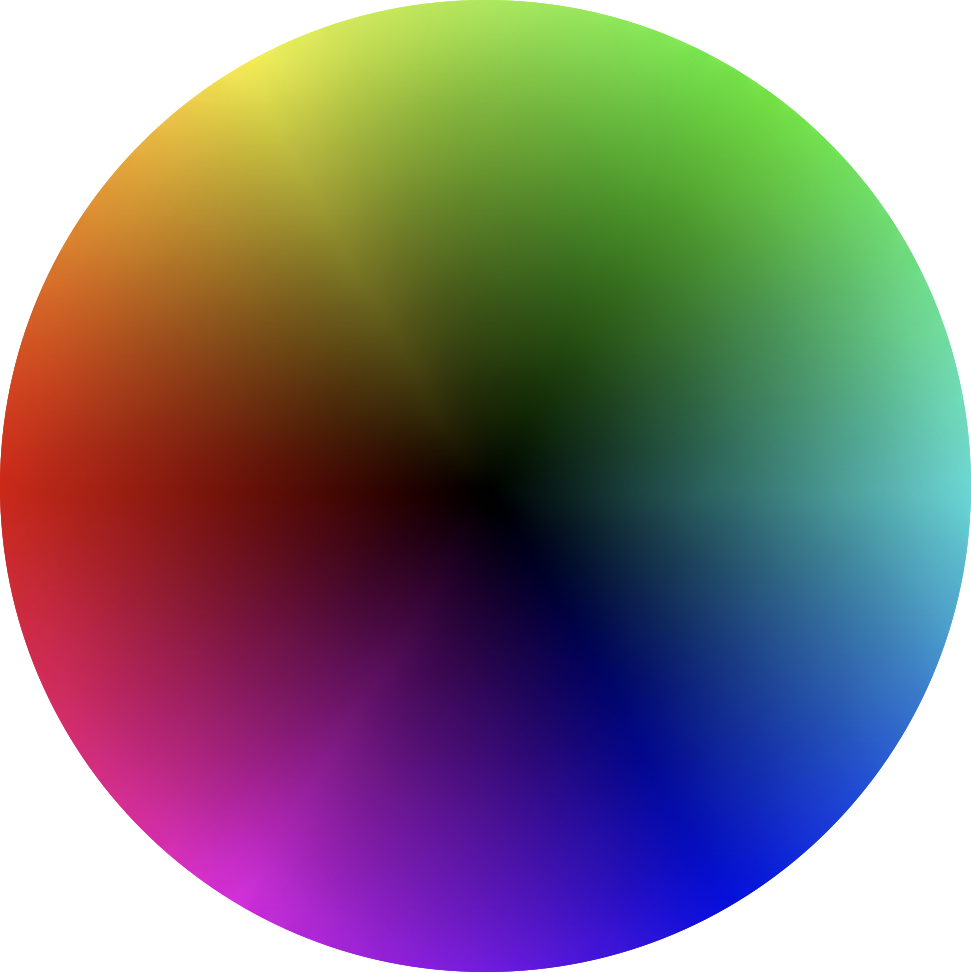}}
        };
      \end{tikzpicture}      
      \fbox{\includegraphics[width=0.23\linewidth]{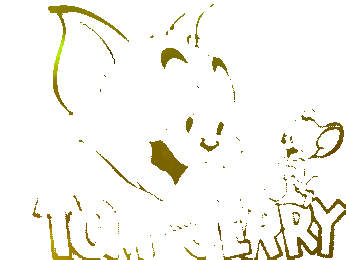}}
      \fbox{\includegraphics[width=0.23\linewidth]{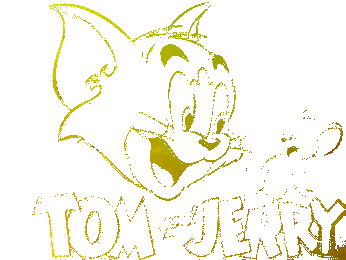}}
      \fbox{\includegraphics[width=0.23\linewidth]{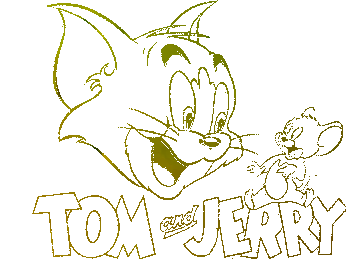}}
    \end{minipage}%
  }

  \caption{Accurate binarization benefits a range of downstream tasks. Left to right: degraded frames, ISauvola~\cite{hadjadj2016isauvola}, Lin~\cite{lin2024neuromorphic}, and Ours. The samples are taken from the HQF and EBT datasets.}
  \label{fig:ex_downstream}
\end{figure}

%% file: src/quan_downstream.tex
\begin{table}[t]
\centering
\caption{Binarization for downstream tasks under varying lighting conditions (\ding{172}~Gentle Light~\ding{173}~Low Light~\ding{174}~Bright Glare), reported as CER~$\downarrow$, RMSE~$\downarrow$, and AEPE~$\downarrow$ respectively.}
\label{tab:quan_downstream}
\resizebox{\columnwidth}{!}{
\begin{tabular}{lc ccc}
\toprule
\multirow{2}{*}{Method} & Dual & OCR & Tracking & Optical Flow \\
\cmidrule(lr){3-3} \cmidrule(lr){4-4} \cmidrule(lr){5-5}
& Modal & \ding{172} / \ding{173} / \ding{174} & \ding{172} / \ding{173} / \ding{174} & \ding{172} / \ding{173} / \ding{174} \\
\midrule
ISauvola~\cite{hadjadj2016isauvola}  &              & 0.62 / 0.82 / 0.79 & 4.25 / 7.72 / 8.91 & 7.62 / 11.85 / 12.01 \\
Bai~\cite{bai2018graph}             &              & 0.54 / 0.73 / 0.71 & 3.41 / 6.84 / 7.68 & 6.24 / 9.76 / 9.44 \\
Zhang~\cite{zhang2024tip}           & $\checkmark$ & 0.38 / 0.48 / 0.49 & 1.72 / 4.92 / 5.62 & 4.15 / 7.82 / 7.56 \\
Lin~\cite{lin2024neuromorphic}      & $\checkmark$ & 0.21 / 0.37 / 0.35 & \best{1.36} / 3.78 / 4.01 & 2.53 / 6.15 / 6.01 \\
\midrule
\textbf{Ours}                       & $\checkmark$ & \best{0.19} / \best{0.29} / \best{0.23} & \best{1.36} / \best{2.41} / \best{2.86} & \best{2.41} / \best{4.92} / \best{4.53} \\
\bottomrule
\end{tabular}
}
\end{table}

%% file: src/ex_framerate.tex
\begin{figure}[!t]
  \centering
  \setlength{\fboxsep}{0pt}
  \setlength{\fboxrule}{0.5pt}
  
  \subfloat[Degradation]{%
    \begin{minipage}{0.23\columnwidth}
      \centering
      \fbox{\includegraphics[trim=0 2cm 0 0, clip, width=\linewidth]{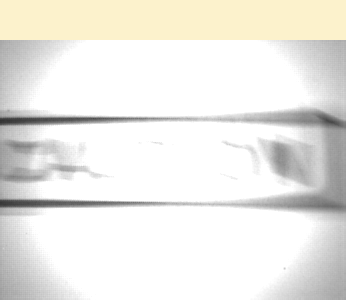}}
    \end{minipage}%
  }\hfill
  \subfloat[500 FPS, f1]{%
    \begin{minipage}{0.23\columnwidth}
      \centering
      \fbox{\includegraphics[trim=0 2cm 0 0, clip, width=\linewidth]{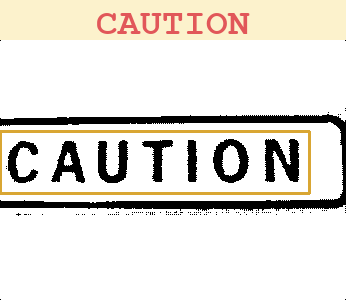}}
    \end{minipage}%
  }\hfill
  \subfloat[1000 FPS, f1]{%
    \begin{minipage}{0.23\columnwidth}
      \centering
      \fbox{\includegraphics[trim=0 2cm 0 0, clip, width=\linewidth]{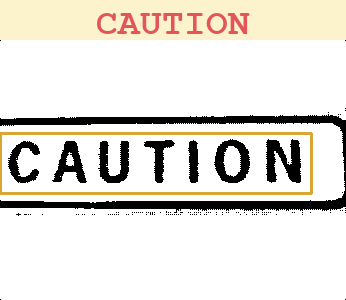}}
    \end{minipage}%
  }\hfill
  \subfloat[5000 FPS, f1]{%
    \begin{minipage}{0.23\columnwidth}
      \centering
      \fbox{\includegraphics[trim=0 2cm 0 0, clip, width=\linewidth]{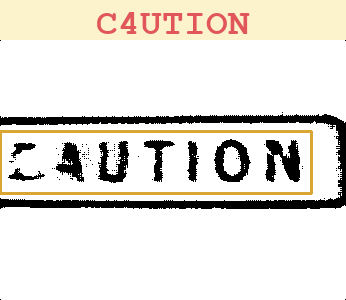}}
    \end{minipage}%
  } \\
  \subfloat[1000 FPS, f2]{%
    \begin{minipage}{0.23\columnwidth}
      \centering
      \fbox{\includegraphics[trim=0 2cm 0 0, clip, width=\linewidth]{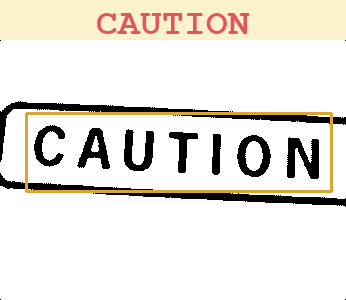}}
    \end{minipage}%
  }\hfill
  \subfloat[5000 FPS, f2]{%
    \begin{minipage}{0.23\columnwidth}
      \centering
      \fbox{\includegraphics[trim=0 2cm 0 0, clip, width=\linewidth]{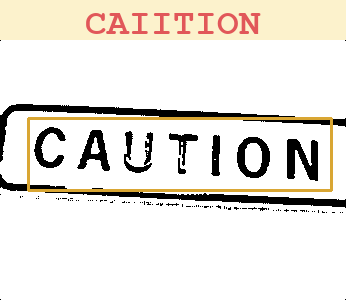}}
    \end{minipage}%
  }\hfill
  \subfloat[1000 FPS, f3]{%
    \begin{minipage}{0.23\columnwidth}
      \centering
      \fbox{\includegraphics[trim=0 2cm 0 0, clip, width=\linewidth]{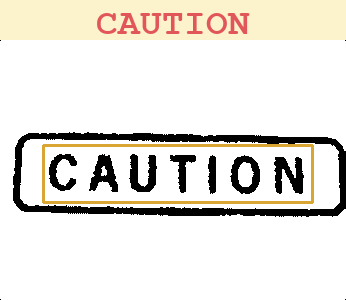}}
    \end{minipage}%
  }\hfill
  \subfloat[5000 FPS, f3]{%
    \begin{minipage}{0.23\columnwidth}
      \centering
      \fbox{\includegraphics[trim=0 2cm 0 0, clip, width=\linewidth]{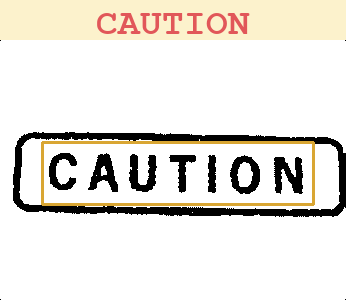}}
    \end{minipage}%
  } 

  \subfloat[MCC]{%
    \begin{minipage}{0.48\columnwidth}
      \centering
      \includegraphics[width=\linewidth]{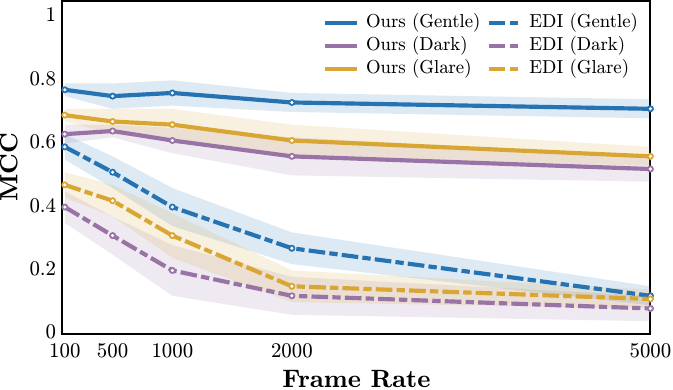}
    \end{minipage}%
  }\hfill
  \subfloat[CER]{%
    \begin{minipage}{0.48\columnwidth}
      \centering
      \includegraphics[width=\linewidth]{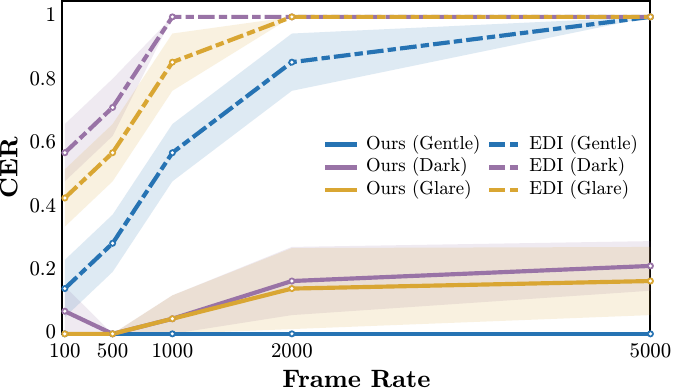}
    \end{minipage}%
  }

  \caption{High-frame-rate binary videos. (b)--(h) present our results for three frames at different time under bright glare, and (i)--(j) provide a comparison under three lighting conditions. The sample is taken from the EBT dataset.}
  \label{fig:ex_framerate}
\end{figure}

%% file: src/ex_runtime.tex
\begin{figure}[!t]
  \centering
  \subfloat[Runtime]{%
    \begin{minipage}{0.48\columnwidth}
      \includegraphics[width=\linewidth]{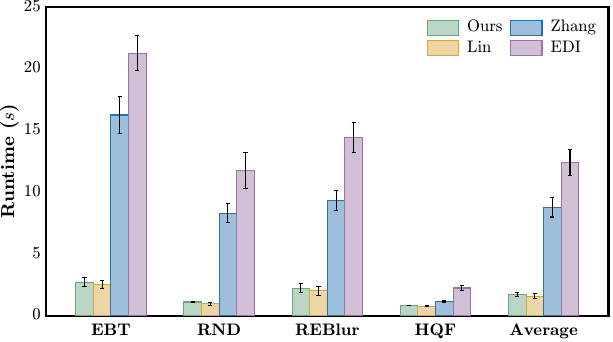}
    \end{minipage}%
  }\hfill 
  \subfloat[Real-Time Factor]{%
    \begin{minipage}{0.48\columnwidth}
      \includegraphics[width=\linewidth]{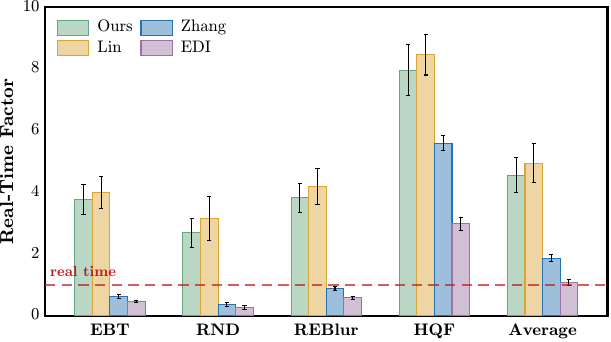}
    \end{minipage}%
  }\\
  \subfloat[Processing Rate]{%
    \begin{minipage}{0.48\columnwidth}
      \includegraphics[width=\linewidth]{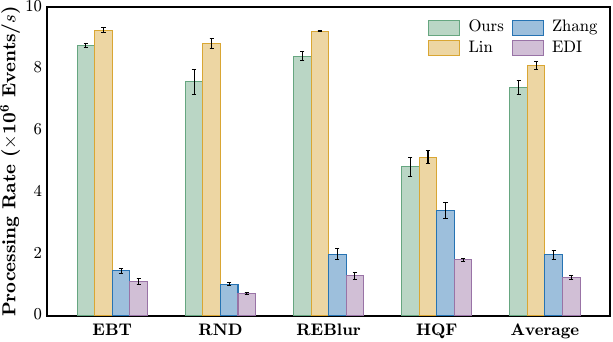}
    \end{minipage}%
  }\hfill
  \subfloat[Processing Latency]{%
    \begin{minipage}{0.48\columnwidth}
      \includegraphics[width=\linewidth]{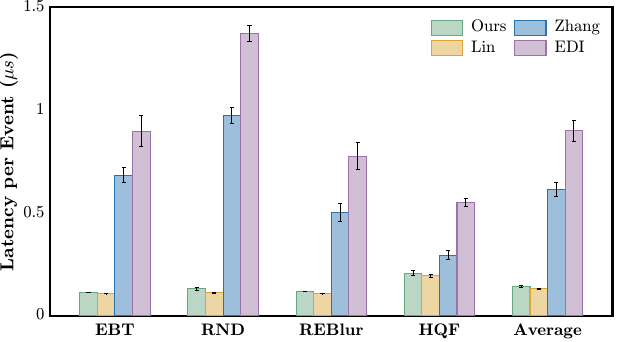}
    \end{minipage}%
  }\\[2pt] 

  \caption{Runtime performance across the samples with varying event rates.}
  \label{fig:ex_runtime}
\end{figure}

%% file: src/ex_ablation.tex
\begin{figure}[!t]
  \centering
  \setlength{\fboxsep}{0pt}
  \setlength{\fboxrule}{0.5pt}
  
  \subfloat[Left to right: degraded frame, $c \uparrow$, $c \downarrow$, and adaptive $c$]{%
    \begin{minipage}{\columnwidth}
      \centering
      \fbox{\includegraphics[width=0.23\linewidth]{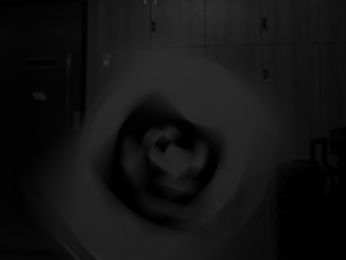}}
      \fbox{\includegraphics[width=0.23\linewidth]{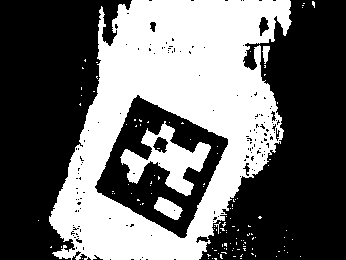}}
      \fbox{\includegraphics[width=0.23\linewidth]{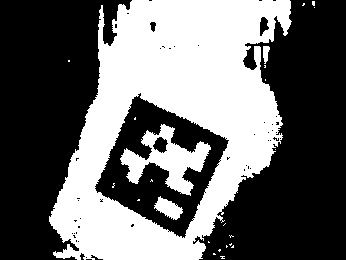}}
      \fbox{\includegraphics[width=0.23\linewidth]{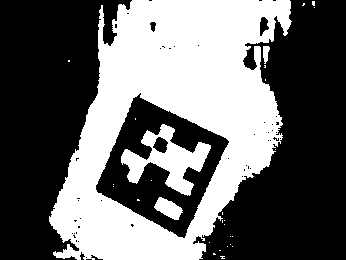}}
    \end{minipage}%
  } \\[6pt] 

  \subfloat[Left to right: degraded frame, $\lambda=1$, $\lambda=0.01$, and adaptive $\lambda$]{%
    \begin{minipage}{\columnwidth}
      \centering
      \fbox{\includegraphics[width=0.23\linewidth]{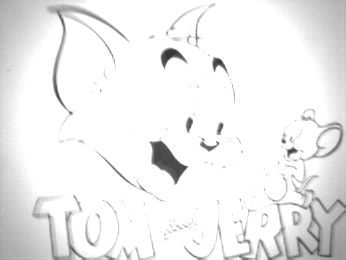}}
      \fbox{\includegraphics[width=0.23\linewidth]{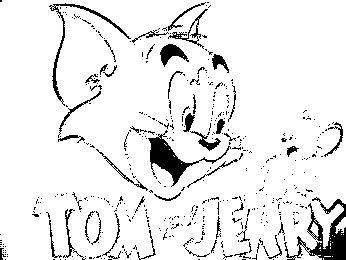}}
      \fbox{\includegraphics[width=0.23\linewidth]{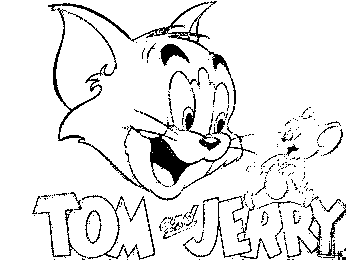}}
      \fbox{\includegraphics[width=0.23\linewidth]{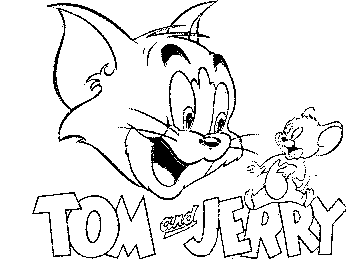}}
    \end{minipage}%
  } \\[6pt] 

  \subfloat[Dark and glare (left), w/o (middle two), w/ (right) structural priors]{%
    \begin{minipage}{\columnwidth}
      \centering
      \fbox{\includegraphics[width=0.23\linewidth]{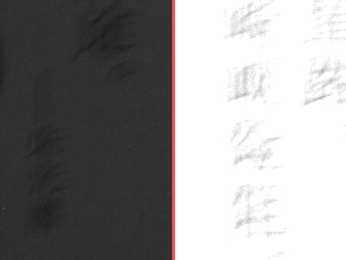}}
      \fbox{\includegraphics[width=0.23\linewidth]{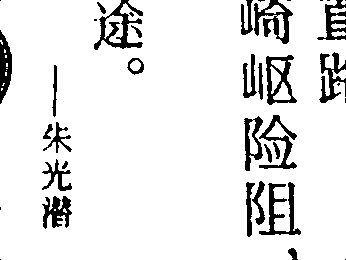}}
      \fbox{\includegraphics[width=0.23\linewidth]{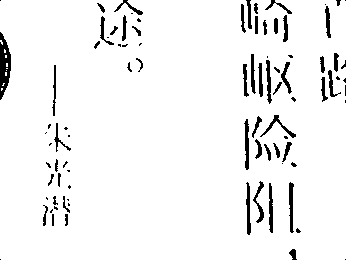}}
      \fbox{\includegraphics[width=0.23\linewidth]{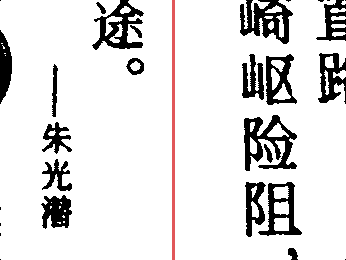}}
    \end{minipage}%
  }

  \caption{Adaptive parameters optimize binarization results.}
  \label{fig:ex_ablation}
\end{figure}

%% file: src/quan_ablation.tex
\begin{table}[t]
\centering
\caption{Evaluations of parameter settings on the EBT dataset (with the \best{best}), reported as \ding{172}~MCC~$\uparrow$~\ding{173}~PSNR~$\uparrow$~\ding{174}~NRM~$\downarrow$.}
\label{tab:quan_ablation}
\resizebox{\columnwidth}{!}{
\begin{tabular}{cc ccc}
\toprule
\multirow{2}{*}{Parameter} & \multirow{2}{*}{Value} & Gentle Light & Low Light & Bright Glare \\
\cmidrule(lr){3-3} \cmidrule(lr){4-4} \cmidrule(lr){5-5}
 & & \ding{172} / \ding{173} / \ding{174} & \ding{172} / \ding{173} / \ding{174} & \ding{172} / \ding{173} / \ding{174} \\
\midrule

\multirow{3}{*}{$c$} 
 & $\uparrow$   & 0.71 / 13.33 / 0.22 & 0.48 / 6.84 / 0.31 & 0.62 / 10.09 / 0.24 \\
 & $\downarrow$ & 0.70 / 13.42 / 0.22 & 0.45 / 7.01 / 0.33 & 0.65 / 10.39 / 0.23 \\
 & Adaptive     & \best{0.73} / \best{13.67} / \best{0.18} & \best{0.54} / \best{7.32} / \best{0.29} & \best{0.66} / \best{10.53} / \best{0.22} \\
\midrule

\multirow{3}{*}{$\lambda$} 
 & 1            & 0.65 / 13.02 / 0.21 & 0.45 / 6.80 / 0.32 & 0.60 / 10.35 / 0.26 \\
 & 0.01            & 0.68 / 12.87 / 0.20 & 0.51 / 6.89 / 0.32 & 0.64 / 10.06 / 0.23 \\
 & Adaptive     & \best{0.73} / \best{13.67} / \best{0.18} & \best{0.54} / \best{7.32} / \best{0.29} & \best{0.66} / \best{10.53} / \best{0.22} \\
\midrule

\multirow{2}{*}{$g(\theta)$} 
 & 0$^\dagger$ & 0.71 / 13.33 / 0.20 & 0.51 / 7.14 / 0.30 & 0.60 / 10.19 / 0.27 \\
 & Adaptive       & \best{0.73} / \best{13.67} / \best{0.18} & \best{0.54} / \best{7.32} / \best{0.29} & \best{0.66} / \best{10.53} / \best{0.22} \\
\bottomrule
\end{tabular}
}
\\[4pt]
\raggedright \footnotesize $\dagger$ Eq.~\eqref{eq:thes_est} is spatially blind to structural cues when $g(\theta)=0$.
\end{table}

%% file: src/ex_quality.tex
\begin{figure}[!t]
  \centering
  \setlength{\fboxsep}{0pt}
  \setlength{\fboxrule}{0.5pt}
  
  \subfloat[Left to right: blurry frame, events with real-world noise, Zhang~\cite{zhang2024tip}, and Ours]{%
    \begin{minipage}{\columnwidth}
      \centering
      \fbox{\includegraphics[width=0.23\linewidth]{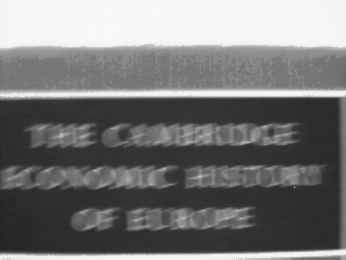}}
      \fbox{\includegraphics[width=0.23\linewidth]{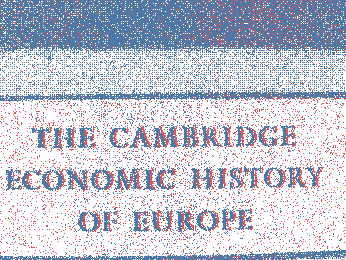}}
      \begin{tikzpicture}[baseline=(main1.base)] 
        \node[inner sep=0pt] (main1) {
        \fbox{\includegraphics[width=0.23\linewidth]{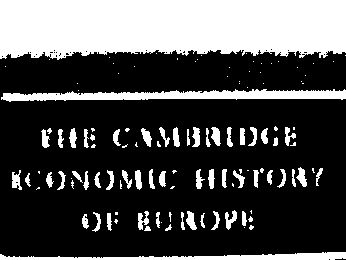}}
        };
        \node[inner sep=0pt, anchor=north east] at (main1.north east) {
            \fcolorbox{boxblue}{white}{\includegraphics[width=0.09\linewidth]{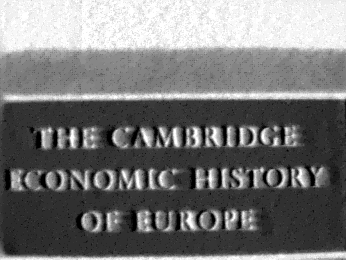}}
        };
      \end{tikzpicture}    
      \fbox{\includegraphics[width=0.23\linewidth]{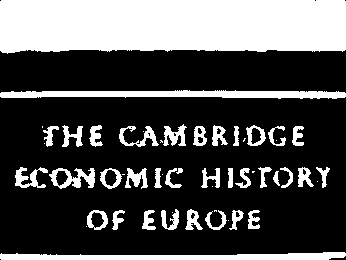}}
    \end{minipage}%
  } \\[6pt] 

  \subfloat[Left to right: blurry frame, sparse events, Ours, and Ours at another time]{%
    \begin{minipage}{\columnwidth}
      \centering
      \fbox{\includegraphics[width=0.23\linewidth]{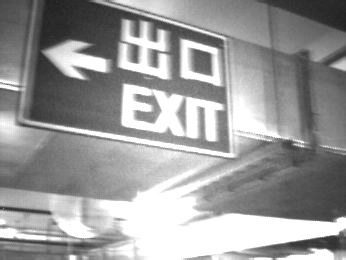}}
      \fbox{\includegraphics[width=0.23\linewidth]{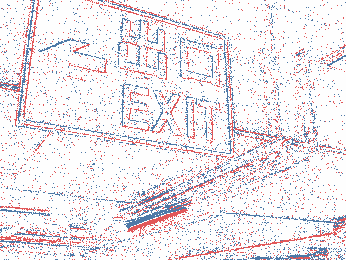}}
      \fbox{\includegraphics[width=0.23\linewidth]{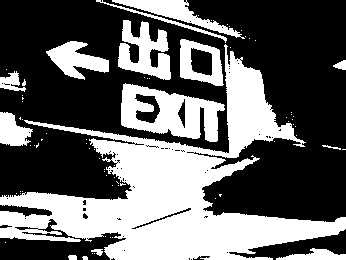}}
      \begin{tikzpicture}[baseline=(main1.base)] 
        \node[inner sep=0pt] (main1) {
        \fbox{\includegraphics[width=0.23\linewidth]{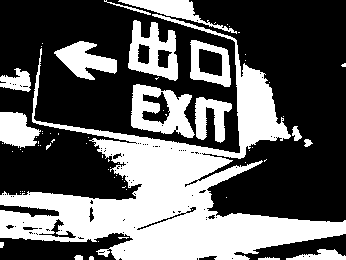}}
        };
        \node[inner sep=0pt, anchor=south east] at (main1.south east) {
            \fbox{\includegraphics[width=0.09\linewidth]{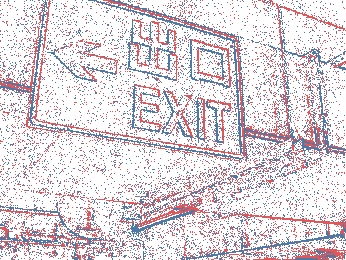}}
        };
      \end{tikzpicture}  
    \end{minipage}%
  }

  \subfloat[NRM]{%
    \begin{minipage}{0.48\columnwidth}
      \centering
      \includegraphics[width=\linewidth]{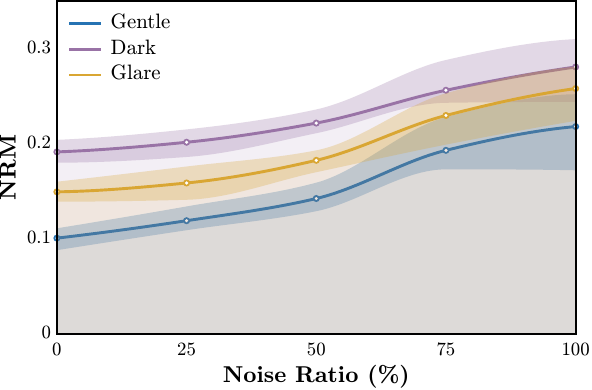}
    \end{minipage}%
  }\hfill
  \subfloat[MCC]{%
    \begin{minipage}{0.48\columnwidth}
      \centering
      \includegraphics[width=\linewidth]{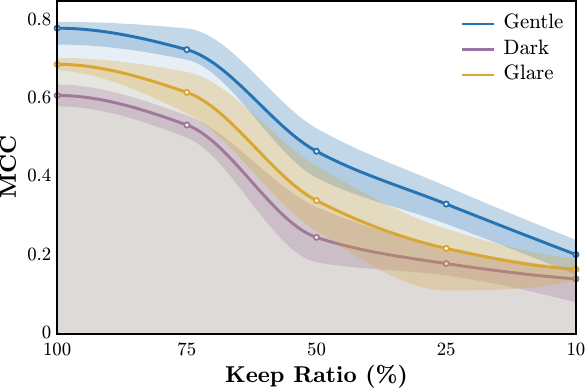}
    \end{minipage}%
  }

  \caption{Evaluation of the impact of event quality on binarization. The samples are taken from the RND and EBT datasets.}
  \label{fig:ex_quality}
\end{figure}

%% file: src/deployment.tex
\begin{table}[t]
\centering
\caption{Edge deployment performance at \ding{172}~$100$~\ding{173}~$1000$~\ding{174}~$2000$ FPS.}
\label{tab:deployment}
\resizebox{\columnwidth}{!}{
\begin{tabular}{lccc}
\toprule
\multirow{2}{*}{Metric} & EDI~\cite{pan2019bringing} & CF~\cite{scheerlinck2018continuous} & Ours \\
\cmidrule(lr){2-2} \cmidrule(lr){3-3} \cmidrule(lr){4-4}
 & \ding{172} / \ding{173} / \ding{174} & \ding{172} / \ding{173} / \ding{174} & \ding{172} / \ding{173} / \ding{174} \\
\midrule
MCC
& 0.58 / 0.40 / 0.28
& 0.54 / 0.54 / 0.51
& 0.69 / 0.67 / 0.63 \\

Runtime (s)
& 18.4 / 58.7 / 115.5 \best{(0.12)}
& 6.8 / 7.2 / 7.3 \best{(1.92)}
& 5.4 / 5.6 / 6.1 \best{(2.3)} \\

Memory (MB)
& 394 / 429 / 453 \second{(1.15$\times$)}
& 308 / 321 / 328 \second{(1.06$\times$)}
& 242 / 246 / 254 \second{(1.05$\times$)} \\

\bottomrule
\end{tabular}
}
\\[4pt]
\begin{flushleft}
\footnotesize
\begin{itemize}[leftmargin=*, labelsep=0.5em]
    \item Runtime excludes disk I/O.
    \item Memory denotes peak working memory excluding the output-frame buffer.
    \item We highlight the \best{real-time factor} at $2000$ FPS, and \second{memory growth} from $100$ to $2000$ FPS.
\end{itemize}
\end{flushleft}
\end{table}

%% file: src/complexity.tex
\begin{table}[t]
\centering
\caption{Computational complexity analysis.}
\label{tab:complexity}
\resizebox{\columnwidth}{!}{
\begin{tabular}{lrr}
\toprule
 & Time & Space \\
\midrule
Spatial decomposition 
& $\mathcal{O}(N)$ 
& $\mathcal{O}(N)$ \\

Contrast estimation 
& $\mathcal{O}(K+|\mathbb{D}|)$ 
& $\mathcal{O}(1)$ \\

Dynamic range reshaping 
& $\mathcal{O}(N)$ 
& $\mathcal{O}(N)$ \\

Threshold estimation 
& $\mathcal{O}(N+N_b)$ 
& $\mathcal{O}(N_b)$ \\

Asynchronous state propagation 
& $\mathcal{O}(K)$ 
& $\mathcal{O}(N)$ \\

Local consistency check
& $\mathcal{O}(|\mathcal{W}|K_\mathrm{flip})$ 
& $\mathcal{O}(1)$ \\

\midrule
Initial binary estimation 
& $\mathcal{O}(K+N+N_b)$ 
& $\mathcal{O}(N+N_b)$ \\

High-frame-rate propagation 
& $\mathcal{O}(K+|\mathcal{W}|K_\mathrm{flip})$ 
& $\mathcal{O}(N)$ \\
\bottomrule
\end{tabular}
}
\\[4pt]
\begin{flushleft}
\footnotesize
\begin{itemize}[leftmargin=*, labelsep=0.5em]
    \item $N_b$ is the number of quantized intensity bins.
    \item $K_\mathrm{flip}$ is the number of triggered binary flips.
    \item $|\mathcal{W}|=9$ for a $3\times3$ local window, and $K_{\mathrm{flip}}\le K$ such that high-frame-rate propagation remains linear.
\end{itemize}
\end{flushleft}
\end{table}